\documentclass{aa}

\usepackage{graphicx}
\usepackage{txfonts}
\usepackage{xcolor}
\usepackage{soul}
\usepackage{hyperref}

\renewcommand\linenumbers{}
\begin{document}

\title{The distribution and origin of metals \\ in simulated Milky Way-like galaxies}

\author{
    F. G. Iza\inst{1,2} \and
    C. Scannapieco\inst{2} \and
    S. E. Nuza\inst{1} \and
    R. Pakmor\inst{3} \and
    R. J. J. Grand\inst{4,5} \and
    \\ F. A. G\'omez\inst{6} \and
    V. Springel\inst{3} \and
    F. Marinacci\inst{7,8} \and
    F. Fragkoudi\inst{9}
}

\institute{
    Instituto de Astronom\'ia y F\'isica del Espacio (CONICET-UBA), 1428 Buenos Aires, Argentina \\
    \email{fiza@iafe.uba.ar} \and
    Universidad de Buenos Aires, Facultad de Ciencias Exactas y Naturales, Departamento de Física. Buenos Aires, Argentina \and
    Max-Planck-Institut für Astrophysik, Karl-Schwarzschild-Str 1, D-85748 Garching, Germany \and
    Instituto de Astrof\'isica de Canarias, Calle V\'ia L\'actea s/n, E-38205 La Laguna, Tenerife, Spain \and
    Departamento de Astrofísica, Universidad de La Laguna, Av. del Astrofísico Francisco Sánchez s/n, E-38206 La Laguna, Tenerife, Spain \and
    Departamento de Astronom\'ia, Universidad de La Serena, Avenida Juan Cisternas 1200, La Serena, Chile \and
    Department of Physics \& Astronomy ``Augusto Righi'', University of Bologna, via Gobetti 93/2, I-40129 Bologna, Italy \and
    INAF, Astrophysics and Space Science Observatory Bologna, Via P. Gobetti 93/3, I-40129 Bologna, Italy \and
    Institute for Computational Cosmology, Department of Physics, Durham University, South Road, Durham DH1 3LE, United Kingdom
}

\date{Received 27 March 2025; accepted 23 July 2025}

\abstract
{
Chemical properties of stellar populations are a key observable that can be used to shed light on the assembly history of galaxies across cosmic time.
In this study, we investigate the distribution and origin of chemical elements in different stellar components of simulated Milky Way-like galaxies in relation to their mass assembly history, stellar age, and metallicity.
Using a sample of 23 simulated galaxies from the Auriga project, we analysed the evolution of heavy elements produced by stellar nucleosynthesis.
To study the chemical evolution of the stellar halo, bulge, and warm (thick) and cold (thin) discs of the model galaxies, we applied a decomposition method to characterise the distribution of chemical elements at $z=0$ and traced back their origin.
Our findings indicate that each stellar component has a distinctive chemical trend despite galaxy-to-galaxy variations.
Specifically, stellar haloes are $\alpha$-enhanced relative to other components, representing the oldest populations, with $\mathrm{[Fe/H]} \sim -0.6$ and a high fraction of ex situ stars of $\sim50$\%. They are followed by the warm ($\mathrm{[Fe/H]} \sim -0.1$) and cold ($\mathrm{[Fe/H]} \sim 0$) discs, with in situ fractions of $\sim90$\% and $\sim95$\%, respectively.
Alternatively, bulges are mainly formed in situ but host more diverse stellar populations, with [Fe/H] abundance extending over $\sim1~\mathrm{dex}$ around the solar value.
We conclude that one of the main drivers shaping the chemical properties of the galactic components in our simulations is the age-metallicity relation.
The bulges are the least homogeneous component of the sample, as they present different levels of contribution from young stars in addition to the old stellar component.
Conversely, the cold discs appear very similar in all chemical properties, despite important differences in their typical formation times.
Finally, we find that a significant fraction of stars in the warm discs were in the cold disc component at birth. We discuss the possible connections of this behaviour with the development of bars and interactions with satellites.
}


\maketitle

\section{Introduction}

Galaxies form at the centres of dark matter haloes, which emerge from the amplification of small density perturbations in the early Universe.
These haloes evolve in a hierarchical fashion, growing through the smooth accretion of material from the intergalactic medium and interactions such as collisions and mergers with smaller structures.
Initially, galaxies consist predominantly of hydrogen and helium, the primordial elements synthesised during Big Bang nucleosynthesis.
However, as stellar populations evolve and complete their life cycles, they produce and expel heavier elements into the interstellar medium (ISM) over diverse timescales.
Mergers further influence the chemical enrichment of galaxies by introducing additional material with distinct metal content and by enhancing the star formation rate of the host.

During the past decades, detailed and precise chemical data of the Milky Way (MW) and external galaxies have enabled the study of the abundances of the different galactic components and their evolution \citep[see e.g.][for a recent review on the galactic chemical evolution of the MW]{Mateucci21}.
In particular, the Gaia survey, with its unprecedented massive database of more than a billion stars, has revolutionised our understanding of the MW \citep{Gaia2018}.
Using the chemical composition of stars as a fossil record of their formation environments, it is possible to gain insight into the processes that shaped galaxies over cosmic time.
For instance, from the observed chemical and kinematic patterns of a variety of stellar populations in the MW, different aspects on its formation and evolution have been addressed, including its star formation history \citep{Snaith2015, Xiang2022}; the chemical patterns of the different galactic components \citep{Schuster2012, An2015, Hayden2017, Barbuy2018, Hayden2020, Bensby2021, Bird2021}; the existence of previous merger events \citep{Helmi18, Helmi2020}; and the discovery of fossil structures such as a primeval thin disc formed in the first billion years of cosmic history \citep{Nepal24}, among many other studies.

Understanding the formation and evolution of different stellar components is crucial to elucidating the processes behind galaxy formation.
Generally speaking, galactic components can be divided into two broad categories based on the kinematics of their stellar populations: \emph{i}) spheroids, dominated by velocity dispersion, such as the bulge and stellar halo, and \emph{ii}) discs, consisting of rotationally supported flattened structures often split into thin (`cold') and thick (`warm') components.
In the MW, the bulge and stellar halo are found to consist of old stellar populations with metallicities in the range $-1.5 \lesssim \mathrm{[Fe/H]} \lesssim 0.5$ \citep{Barbuy2018} and $-4 \lesssim \mathrm{[Fe/H]} \lesssim 0$ \citep{Ryan91, Schorck09}, respectively.
Conversely, disc components are mainly formed by younger stellar populations with higher median metallicities that are close to the solar value, showing a dichotomy in the [O/Fe] abundance, with the thin disc having lower values in comparison to the thick disc \citep[see e.g.][and references therein]{Palla20}.
However, an increasing dependence of the low-$\alpha$/high-$\alpha$ ratio with Galactocentric distance has been observed \citep{Anders14}, suggesting that the MW thick disc is concentrated in the inner regions.

Several mechanisms have been proposed to explain the formation of bulges, including the accretion of stellar systems onto the galactic centre by dynamical friction, gas accretion during mergers, metal-rich inflows from the thick disc or halo, and the accumulation of material towards central regions simply reflecting the initial conditions of galaxy formation \citep[e.g.][]{WG92, Gargiulo2019}.
In particular, chemical evolution models have established that the metal-poor population of bulge stars most likely formed on very short timescales, thus favouring an in situ formation scenario \citep{Matteucci19}, whereas metal-rich stars could have been accreted directly from the thin or thick discs, explaining the observed high-metallicity tails \citep{Debattista17}.

Regarding the assembly of the stellar halo, proposed formation channels include in situ scenarios that consider an interplay between gas inflows from the galactic halo and outflows preventing star formation \citep{Hartwick76, Prantzos03, Brusadin13} as well as ex situ scenarios involving the accretion of stars from galaxy mergers \citep{Monachesi2019, MB20}.
In particular, it has been shown that the MW inner halo is highly contaminated with stars that belonged to the so-called Gaia-Enceladus-Sausage system that presumably merged with our Galaxy about $10~\mathrm{Gyr}$ ago \citep{Helmi18, Belokurov18}.
\cite{Helmi18} concluded that this encounter was crucial for the build-up of the MW stellar halo and also led to the dynamical heating of the precursor of the thick disc, thus contributing to its formation.
\cite{MB20} estimated that $30$--$50\%$ of the stellar halo mass in the MW corresponds to stars acquired during the Gaia-Enceladus event, concluding that most of the stellar mass in the Galactic halo was accreted.

\begin{table*}
	\caption{Galactic properties at $z=0$.}
	\centering
	\label{tab:galactic_properties}
	\begin{tabular}{lccccccc}
		\noalign{\smallskip} \hline \noalign{\smallskip}
		Galaxy   & $R_{200}$ & $M_{200}$ & $M_\star$ & $M_\mathrm{gas}$ & D/T & $R_\mathrm{d}$ \\
                     & [kpc] & [$10^{10}~\mathrm{M}_\odot$] & [$10^{10}~\mathrm{M}_\odot$] & [$10^{10}~\mathrm{M}_\odot$] & & [kpc] \\
		\noalign{\smallskip} \hline \noalign{\smallskip}
		Au2               & 261.7 & 191.4 & 9.4   & 12.5  & 0.81  & 33.7  \\
		Au3               & 239.0 & 145.8 & 8.7   & 9.7   & 0.74  & 23.6  \\
		Au4               & 236.3 & 140.9 & 8.8   & 12.8  & 0.38  & 21.8  \\
		Au6               & 212.8 & 102.9 & 5.6   & 6.4   & 0.79  & 18.2  \\
		Au7               & 218.9 & 112.0 & 5.9   & 11.6  & 0.60  & 21.7  \\
		Au8               & 216.3 & 108.1 & 4.0   & 9.5   & 0.84  & 29.2  \\
		Au9               & 215.8 & 107.3 & 7.5   & 6.5   & 0.72  & 9.9   \\
		Au10              & 214.0 & 104.7 & 6.2   & 8.6   & 0.73  & 8.4	  \\
		Au12              & 217.1 & 109.3 & 6.6   & 8.8   & 0.68  & 14.9  \\
		Au13              & 223.2 & 118.8 & 6.3   & 10.6  & 0.84  & 13.0  \\
		Au14              & 249.4 & 165.7 & 11.7  & 14.1  & 0.61  & 17.7  \\
		Au15              & 225.4 & 122.2 & 4.3   & 9.3   & 0.69  & 19.2  \\
		Au16              & 241.4 & 150.3 & 7.0   & 10.4  & 0.88  & 31.1  \\
		Au17              & 215.7 & 107.1 & 9.0   & 7.6   & 0.76  & 15.0  \\
		Au18              & 225.3 & 122.1 & 8.4   & 7.1   & 0.80  & 14.0  \\
		Au20              & 227.0 & 124.9 & 5.6   & 14.2  & 0.72  & 24.1  \\
		Au21              & 238.6 & 145.1 & 8.7   & 11.8  & 0.80  & 18.6  \\
		Au22              & 205.5 & 92.6  & 6.2   & 3.6   & 0.69  & 7.9	  \\
		Au23              & 245.2 & 157.3 & 10.1  & 9.4   & 0.84  & 19.8  \\
		Au24              & 242.7 & 152.5 & 9.3   & 9.7   & 0.74  & 24.1  \\
		Au25              & 225.3 & 122.1 & 3.7   & 7.9   & 0.86  & 22.8  \\
		Au26              & 242.7 & 152.6 & 10.9  & 10.5  & 0.71  & 14.6  \\
		Au27              & 253.8 & 174.5 & 10.3  & 12.5  & 0.71  & 17.2  \\
		\noalign{\smallskip} \hline
	\end{tabular}
    \tablefoot{
    The columns shown are:
        (1) galaxy name,
        (2) virial radius $R_{200}$,
        (3) virial mass $M_{200}$,
        (4) galaxy stellar mass $M_\star$,
        (5) galaxy gas mass $M_\mathrm{gas}$,
        (6) disc-to-total mass fraction and
        (7) disc radius $R_\mathrm{d}$ obtained from \cite{Iza2022}.
    }
\end{table*}

The growth of the thin disc component is commonly described following an `inside-out' formation process \citep{Larson1976}, in which disc galaxies are thought to form the inner regions first and the outer regions at later times.
In particular, this behaviour has been observed in the MW using different stellar populations \citep{Pan2015, Frankel2019, Prantzos2023}.
These observations, however, take into account the positions of stars as they are observed today instead of the position at birth, which is, due to orbital mixing, lost through the evolution.
Since stars retain their chemical composition during their whole lifespan, information regarding the properties of stars at birth may be recovered by detailed analysis of their chemical abundances.
 
As a result of the inside-out scenario, a negative radial gradient is expected to develop in the distribution of stellar ages: older stars are concentrated in the inner regions, and younger stars are in the outer regions of the Galactic disc.
Since stars form from the gas that precedes them, the inner disc regions give rise to more enriched stars, thus creating a negative profile of metal abundances at $z=0$ that has been observed in the MW and other spiral galaxies \citep[e.g.][]{Costa2004, Esteban2005, Rudolph2006, Lemasle2008, Luck2011, Genovali2014, Genovali2015, Balser2015, Magrini2017, Lemasle2018, Stanghellini2018, Palla20, Mateucci21, Gaia2023, Yang2025}.
However, contradictory evidence has also been found regarding the metallicity gradients.
\cite{Stanghellini2010} and \cite{Gibson2013} found relatively flat and temporally invariant abundance profiles using observations of planetary nebulae and cosmological hydrodynamical simulations, respectively.
\cite{Yuan2011}, on the other hand, found steep gradients at high redshift that flatten with time.
The flattening of abundance profiles has also been quantified in simulations \citep{Pilkington2012, Gibson2013, Tissera2016}.
Possible reasons for the flattening of the profile include radial migration, star formation, feedback processes, and mixing due to merger events.

As observations on chemical abundances increase in abundance and detail, cosmological simulations including a treatment of chemical enrichment become a powerful tool to interpret these observations by probing the role of various  physical processes such as accretion, mergers, mixing and migration on the chemical abundances of the ISM and stellar components in galaxies. Since the early models of \cite{Mosconi2001}, \cite{Lia2002}, \cite{Kawata2003}, \cite{Tornatore2004}, \cite{Okamoto2005}, and \cite{Scannapieco2005}, significant progress has been made in the modelling of chemical enrichment, incorporating the chemical pollution via  supernovae Type II and Ia events  and stars in the AGB phase. 
 Chemical enrichment models are currently incorporated in simulations with sophisticated models for star formation and feedback (see \citealt{Naab2017} for a review), which also allows for the inclusion of metal-dependent cooling functions \citep{Wiersma2009a, Wiersma2009b, Vogelsberger2014, Schaye2015} and, in some cases, modelling of the effects of  metal diffusion (e.g. \citealt{Aumer2013}). 

These theoretical efforts have enabled detailed investigations related to the chemical enrichment of Milky Way-like galaxies, such as the origin of the chemical properties of the disc(s) including a possible dichotomy in the abundance of $\alpha$-elements relative to iron and its connection to the formation of the thin and thick discs \citep{Agertz2021, Yu2021, Mackereth2019, Grand2018, Buck2020}, the identification of tracers of accretion events in the stellar halo \citep{Fattahi2019, Buder2024, Monachesi2019, Khoperskov2023}, the chemical properties of the bulge including the effects of a bar component \citep{Tissera2018, Fragkoudi2020, Chen2022}, the origin of metallicity profiles \citep{Tissera2016, Vincenzo2020, Bellardini2022, Sun2025}, and the effects of mergers and gas flows on the overall metallicities of galaxies \citep{Bustamante2018, Grand2019}.

In this work we use the simulations of the Auriga project \citep{Grand2017}, a set of galaxies simulated in high resolution and in a cosmological environment that aim to reproduce galaxies of Milky Way mass.
These simulated galaxies have many properties that resemble those of the actual Milky Way and have been analysed in various previous works (see e.g. \cite{Marinacci2017} for the properties of the H{\scshape i} discs; \cite{Prada2019} for the morphology of the dark matter halo; \cite{Monachesi2019} for the properties of stellar haloes; \cite{Fragkoudi2020} for the chemo-dynamical properties of bars and boxy-peanut bulges; and \cite{Hammer2024} for the accretion history).

When integrated over the disc extent, we found that the accretion rates increase rapidly at early times and show different behaviours for the late evolution.
Accretion rates can decrease, increase, or stay approximately constant in the last $8\,\mathrm{Gyr}$ depending on the galaxy \citep{Iza2022}.
We also found that Milky Way analogues -- in terms of having no significant merger events and suffering no significant instabilities in the recent past -- have in most cases decreasing accretion rates at late times, which translates into a similar behaviour for the star formation rate as a function of time.
In \cite{Iza2024}, we studied the radial dependence of the accretion rates and found that most of the Milky Way analogues are consistent with an inside-out behaviour, where younger (older) stars are preferentially located in the outer (inner) disc regions (see also \citealt{Nuza2019}).
Our results show that gas accretion plays a key role in the determination of the properties of spiral galaxies. 

In this paper, we further explore the Auriga simulations, focusing on the distribution of metals in four galactic components: stellar haloes, bulges, cold discs, and warm discs. We tag each star in a galaxy by using a simple dynamical decomposition method that takes into account the rotation of the star and its location in the potential well generated by the galaxy.
For each component, we analysed the distribution of the iron and $\alpha$-element abundances, the so-called age-metallicity relation, and the metallicity profiles of the cold disc along the galactic plane.
To discuss the origin of the observed abundances, we also studied the origin of these metals by tracking stars from the time of birth up to the present.
In this way, we checked the differences in abundances between in situ stars (stars born in the galaxy) and ex situ stars (born somewhere else) and analysed the component of origin of the stars in a given component today to quantify the component-wise stellar migration inside the galaxy.

The organisation of this paper is as follows:
In Section~\ref{sec:simulations} we describe the main features of the \textsc{arepo} code and the properties of the galaxies of the Auriga project.
In Section~\ref{sec:decomposition} we discuss the dynamical decomposition used to tag the stars of each galaxy as part of the halo, bulge, cold disc, or warm disc.
In Section~\ref{sec:metals} we describe the present-day distribution of metals, focusing on the metal abundances, the age-metallicity relation, and the metallicity profiles.
In Section~\ref{sec:global_trends} we present the global trends found in the distribution of chemical elements.
In Section~\ref{sec:origin} we study the origin of the present-day distribution of metals by tracing the galactic component associated with each star at the time of its formation and comparing it to its current galactic component.
Finally, in Section~\ref{sec:conclusions}, we discuss and summarise our main results.

\section{The Auriga simulations}
\label{sec:simulations}

For this study, we undertook an examination of 23 galaxies drawn from the Auriga project \citep{Grand2017, Grand2024}.
The Auriga project is a collection of high-resolution cosmological simulations carried out using the magnetohydrodynamic (MHD) code \textsc{arepo} \citep{Springel2010}.
\textsc{arepo} is a quasi-Lagrangian, dynamic mesh code designed to track the evolution of MHD and collisionless dynamics within a cosmological framework.
Gravitational forces are computed using a conventional TreePM approach, while the MHD equations are solved through a second-order Runge-Kutta method applied to a dynamic Voronoi mesh.
Our simulations assume the cosmological parameters from \cite{Planck2014} with $\Omega_{\rm M} = 0.307$, $\Omega_{\rm b} = 0.048$, $\Omega_\Lambda = 0.693$, and a Hubble constant of $H_0 = 100 ~ h \, \mathrm{km} \, \mathrm{s}^{-1} \, \mathrm{Mpc}^{-1}$, where $h = 0.6777$.

The galaxy formation model employed in the Auriga project includes processes such as primordial and metal-line cooling, a uniform ultraviolet (UV) background field to account for reionisation, star formation (as per \cite{Springel2003}, with a density threshold of $0.13~\mathrm{cm}^{-3}$ and a star formation timescale of $\tau=2.2~\mathrm{Gyr}$), magnetic fields \citep{Pakmor2014, Pakmor2017, Pakmor2018}, active galactic nuclei, and the incorporation of energetic and chemical feedback from Type II supernovae, as well as mass loss and metal return from Type Ia supernovae and asymptotic giant branch stars \citep{Vogelsberger2013, Marinacci2014, Grand2017}.

\begin{figure}
    \centering
    \includegraphics[draft=false]{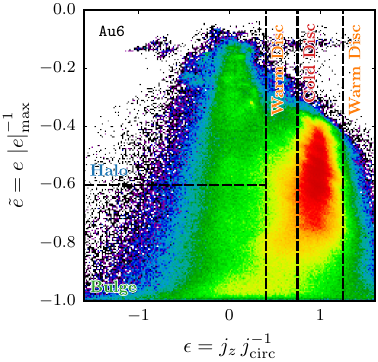}
    \caption{
    Distribution of stars in the gravitational potential-circularity parameter space for Au6 at $z=0$.
    We have normalised the potential to the maximum of its absolute value to make the values independent of galactic mass and restricted them to the range $[-1, 0]$.
    The circularity parameter, on the other hand, has been calculated as $\epsilon = j_z j_\mathrm{circ}^{-1}$, where $j_z$ is the specific angular momentum in the $z$-direction of each star and $j_\mathrm{circ} (R) = R \sqrt{G M_R / R}$ the angular momentum of a circular orbit of radius $R$.
    In this figure, particles near $\epsilon \approx 1$ describe circular orbits, while particles near $\epsilon \approx 0$ show random motion.
    Stars with $\tilde{e} \approx - 1$ are closer to the centre of the galaxy than particles with $\tilde{e} \approx 0$.
    Dashed lines delimit the different components in each panel, as described in the text.
    Fig.~\ref{fig:galaxy_decomposition_phase_space} shows this figure for each galaxy in the selected sample of galaxies.
    }
    \label{fig:phase_space_au6}
\end{figure}

Within the simulation, star particles are used to represent a single stellar population with specific age and metallicity attributes, following a \cite{Chabrier2003} initial mass function.
For single stellar populations, the number of Type Ia supernova events is computed by integrating the delay time distribution function as outlined in \cite{Grand2017}.
The quantities of mass and metals injected into the interstellar medium are subsequently determined using SNIa \citep{Thielemann2003, Travaglio2004} and AGB \citep{Karakas2010} yield tables and are distributed among neighbouring gas cells.
In these simulations, nine chemical elements are tracked: H, He, C, O, N, Ne, Mg, Si, and Fe.

Type II supernova events are assumed to occur instantaneously and are implemented by transforming a stochastically selected star-forming gas cell into either a star or a wind particle.
These particles carry 40\% of the metal mass of the gas cells from which they originate and are given a velocity kick proportional to the local 1D dark matter velocity dispersion before being launched.
Wind particles travel until they reach a cell with a density falling below 5\% of the physical density threshold for star formation or until they reach a maximum travel time corresponding to 2.5\% of the Hubble time at the present time-step.
Upon reaching their destination, the wind particles release their mass, momentum, energy, and metals into the closest gas cell at the time of recoupling.

The mass resolution in our simulations is approximately $3 \times 10^5 ~ \mathrm{M}_\odot$ for dark matter and $5 \times 10^4 ~ \mathrm{M}_\odot$ for baryons.
These values correspond to the level 4 resolution runs as described in \cite{Grand2017}.
The softening length for both star and dark matter particles is held constant in comoving coordinates at $500 ~ h^{-1} \, \mathrm{cpc}$ up to redshift $z=1$.
Beyond this point, the softening length remains fixed at $369 ~ \mathrm{pc}$ in physical units.
For each of the Auriga galaxies, our dataset comprises a total of 128 snapshot files, capturing their entire evolutionary history.
These snapshots are separated on average by approximately $100~\mathrm{Myr}$.

The host haloes housing the Auriga galaxies were selected at redshift $z=0$ from a parent dark matter-only cosmological simulation conducted within a box measuring $100~\mathrm{cMpc}$ on each side \citep{Schaye2015}.
Two main criteria guided the selection of these host haloes.
First, they fell within the virial mass range of 1 to $2 \times 10^{12} ~ \mathrm{M}_\odot$, and second, they were relatively isolated.
This isolation criterion was defined such that the centre of the host halo must be located outside a distance of nine times the virial radius of any other halo with a mass greater than 3\% of the host halo's mass.
We note, however, that this is different from the Local Group environment, which might have an effect on the properties of the Milky Way (see e.g. \citealt{Nuza2014, Creasey2015, Scannapieco2015, Samuel2020, Samuel2021, Biaus2022}).

Our 30 Auriga galaxies, denoted with the `Au' prefix followed by numbers ranging from 1 to 30, were randomly chosen from the most isolated quartile of the host haloes described above.
These selected galaxies exhibit $z=0$ virial masses approximately in the range $9$--$17 \times 10^{11} ~ \mathrm{M}_\odot$, and stellar masses between $3 \times 10^{10}$ and $12 \times 10^{10} ~ \mathrm{M}_\odot$, as summarised in Table~\ref{tab:galactic_properties}.
The virial masses of these galaxies are in agreement with the commonly accepted value for the MW, i.e. approximately $10^{12} ~ \mathrm{M}_\odot$, rendering them `MW-like' or, more specifically, `MW-mass' galaxies. 
The table also provides information on the redshift $z=0$ values of the virial radii ($R_{200}$), which range from approximately $206$ to $262~\mathrm{kpc}$.

Following \cite{Iza2024}, we exclude from this work the galaxies from the Auriga project that were found to consist of strongly perturbed stellar discs.
These galaxies are: Au1, Au5, Au19, Au28, Au29 and Au30.
Furthermore, we also exclude Au11 from the sample because its analysis may prove extremely complex due to an ongoing merger event at $z=0$.

\section{The galactic decomposition}
\label{sec:decomposition}

The objective of this work is to study the present-day distribution of metals in the simulated MW-mass galaxies of the Auriga project, and track back their origin.
It is of particular interest to describe the differences that may arise in different galactic components due to their particular formation history.
Therefore, a proper galactic decomposition is needed in order to correctly identify the different components of each simulated galaxy, at the present day and as a function of time.
In this work, we focus on the `cold' (thin) disc, `warm' (thick) disc, stellar halo, and bulge.
With this in mind, we first diagonalised the inertia tensor of all the stars in the inner $10~\mathrm{ckpc}$ (at each time) and rotate the reference frame such that the principal axis that is closer to the angular momentum vector of the stars lies in the $z$-direction.
The galactic plane therefore coincides with the $xy$ plane at all times.

There are several methods in which the galactic components may be separated.
Since the main goal of this work is to study the distribution of metals, we use only kinematic and gravitational considerations when defining the galactic components.
We then use two stellar parameters: the circularity parameter $\epsilon$ \citep{Scannapieco2009} and the normalised gravitational potential $\tilde{e}$ which we use as a proxy for the galactocentric distance.

\begin{figure*}
    \centering
    \includegraphics[draft=false]{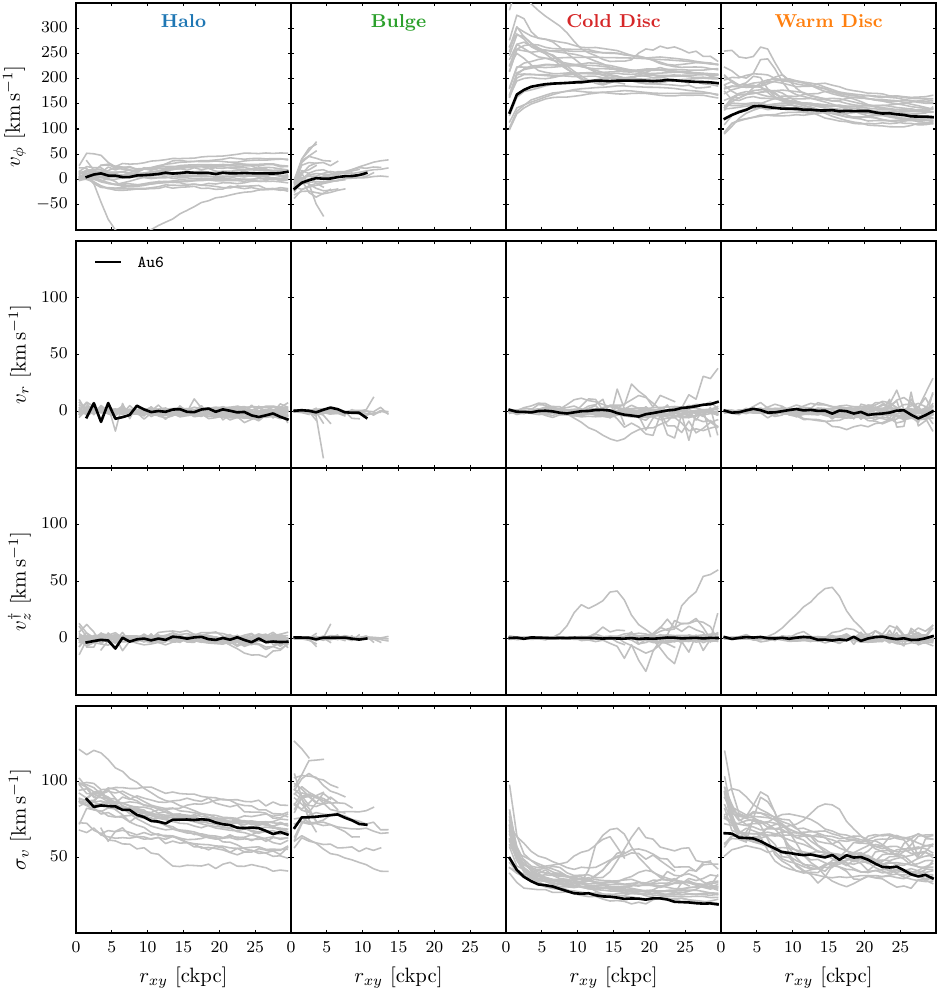}
    \caption{
    Velocity profiles for the stars of the sample by galactic component at $z=0$.
    Each row shows the profiles for a given direction (tangential velocity, radial velocity, and vertical velocity) and the standard deviation (from top to bottom); each column shows the profiles for a given component (halo, bulge, cold disc, and warm disc, from left to right).
    Au6, as it is the reference galaxy, is highlighted in black.
    The vertical velocity is calculated as $v^\dagger_z = v_z \mathrm{sign}(z)$ to distinguish between inflows ($v^\dagger_z < 0$) and outflows ($v^\dagger_z > 0$).
    }
    \label{fig:stellar_velocities_by_region_originals_average}
\end{figure*}

The circularity parameter is defined as the ratio between the angular momentum of a given particle and the angular momentum of a circular orbit of the same radius, $\epsilon = j_z / j_\mathrm{circ}$, where $j_\mathrm{circ} = r\,v_\mathrm{circ}(r)$, with $v_\mathrm{circ}(r) = \sqrt{G M(r) / r}$.
This unconstrained number yields typical values of $\epsilon \approx 0$ for dispersion-dominated stars (which usually populate the spheroid components) and $\epsilon \approx 1$ for rotationally supported stars (which usually populate the discs).

\begin{figure*}
    \centering
    \includegraphics{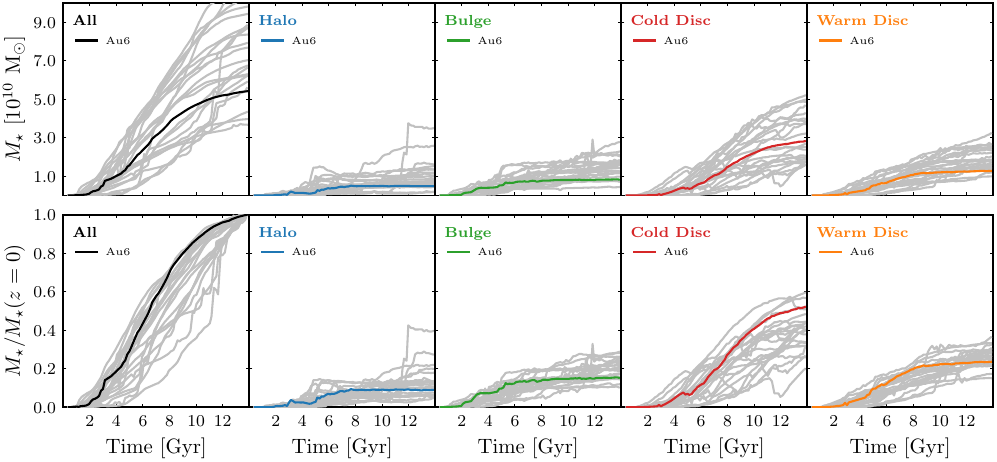}
    \caption{
    Evolution of the stellar mass for the galaxy (first panel) and of each galactic component (subsequent panels).
    Each line corresponds to a galaxy in the sample, highlighting Au6 in colour.
    All lines in the second row are normalised to the present-day galaxy mass.
    }
    \label{fig:decomposition_mass_evolution_normalised}
\end{figure*}

To compute the potential, we first calculated a reference potential for each galaxy using the 16 dark matter particles that are closest to the $3R_{200}$ radius\footnote{This is necessary because the periodic boundary conditions of the simulations are not consistent with a well-defined gravitational potential.} and set this value as the zero-level for the gravitational potential by subtracting it from the potential of each star.
Afterwards, we calculate $\tilde{e}$ for each star, defined as $\tilde{e} = e / \left| e \right|_\mathrm{max}$, where $e$ is the gravitational potential of each star and $\left| e \right|_\mathrm{max}$ the absolute value of the gravitational potential of the star closest to the centre of the potential well.
In this way, $\tilde{e}$ is constrained between $-1$ and 0, and indicates how close each star is to the centre of the galaxy ($\tilde{e} \approx -1$ corresponds to stars near the centre of the potential well, $\tilde{e} \approx 0$ corresponds to stars that are far from the centre of the halo)\footnote{This approach is similar to that of \cite{Obreja2018} and \cite{Du2020}, although they employ the binding energy instead of the gravitational potential and a Gaussian Mixture Model.
Notably, applying the same decomposition procedure using the total energy of each star instead of the gravitational potential does not yield significant differences in the conclusions drawn throughout this work.
}.
Additionally, normalising the potential to the maximum absolute value ensures that the parameter is independent of halo mass.

By using these two parameters, we can separate rotating and non-rotating components, and components closer and farther away from the centre of the halo.
In order to show how we perform the galactic decomposition, we use Au6 as an example galaxy, as its properties lie near the average quantities of the whole sample.
Fig.~\ref{fig:phase_space_au6} shows the distribution of stars in the $\epsilon$-$\tilde{e}$ plane as a heat map for Au6.
From this phase space, we defined four parameters to assign stars to each galactic component:

\begin{itemize}
    \item $\epsilon_\mathrm{d}$ is the circularity parameter of the star particles following circular orbits.
    We adopted $\epsilon_\mathrm{d} = 1$.
    \item $\epsilon_\mathrm{rot}$ is the threshold we used to separate rotating components (discs) from non-rotating components (bulge and halo).
    We adopted $\epsilon_\mathrm{rot} = 0.4$.
    \item $\delta_\epsilon$ is the allowed width of the cold (thin) disc particles around $\epsilon_\mathrm{d}$.
    We adopted $\delta_\epsilon = 0.25$.
    \item $\tilde{e}_0$ is the threshold between the component closest to the centre of the galaxy (bulge) and the component farthest from from the centre of the galaxy (halo).
    We adopted $\tilde{e}_0 = -0.6$.
\end{itemize}

Galactic components were then defined by the following prescription:

\begin{itemize}
    \item The halo consists of non-rotating ($\epsilon < \epsilon_\mathrm{rot}$) stars located farther from the centre of the galaxy ($\tilde{e} > \tilde{e}_0$).
    \item The bulge consists of non-rotating ($\epsilon < \epsilon_\mathrm{rot}$) stars located closer to the centre of the galaxy ($\tilde{e} \leq \tilde{e}_0$).
    \item The cold disc consists of rotating stars ($\epsilon \geq \epsilon_\mathrm{rot}$) with a low velocity dispersion ($\left| \epsilon - \epsilon_\mathrm{d} \right| \leq \delta_\epsilon$).
    \item The warm disc consists of rotating stars ($\epsilon \geq \epsilon_\mathrm{rot}$) with a high velocity dispersion ($\left| \epsilon - \epsilon_\mathrm{d} \right| > \delta_\epsilon$).
\end{itemize}

The separation between the different components is included in Fig.~\ref{fig:phase_space_au6} for Au6, from where the presence of cold and hot discs, a non-rotating bulge and a low mass stellar halo are evident.
These four components are also present in the rest of the Auriga sample (as shown in Fig.~\ref{fig:galaxy_decomposition_phase_space}) although with the expected galaxy-to-galaxy variations.

It is worth noting that the adopted values for $\epsilon_\mathrm{d}$, $\epsilon_\mathrm{rot}$, $\delta_\epsilon$, and $\tilde{e}_\mathrm{0}$ to separate components are adequate for all galaxies in the sample.
In fact, our choice for the assumed parameters is done to ensure that the decomposition properly describes the different dynamical properties of the four components, as we show in Fig.~\ref{fig:stellar_velocities_by_region_originals_average}, where the radial, azimuthal and vertical velocity profiles of the four components of Au6 (black lines) and the rest of the galaxies (grey lines) are exhibited, along with the total velocity dispersion.
As expected, our decomposition yields near zero velocities (in all directions) with high velocity dispersion for the spheroidal components.
The cold discs show azimuthal velocities in the range $150$--$300~\mathrm{km}\,\mathrm{s}^{-1}$ with low velocity dispersion ($0$--$50~\mathrm{km}\,\mathrm{s}^{-1}$), while the warm discs show, comparatively, smaller velocities and higher dispersions.
It is also worth mentioning that the bulges extend typically up to about $5$--$10~\mathrm{kpc}$, with some variations among galaxies.
Additionally, the outliers observed in the figure -- negative azimuthal velocities in the halo and positive bumps in the vertical velocities of the discs -- are the result of the interaction with orbiting satellites in the cases of Au14, Au17 and Au20.

Most galaxies in the Auriga sample are, at $z=0$, rotationally supported, disc-dominated systems, with minor populations orbiting the galactic centre in random motion as part of the spheroids (Fig.~\ref{fig:galaxy_decomposition_phase_space}).
This is linked to the selection criteria: Auriga galaxies generally exhibit relatively quiescent formation histories, with most experiencing no significant mergers in the recent past (see e.g. \citealt{Grand2017}; \citealt{Iza2022}; and Section~\ref{sec:origin})
However, as shown in Fig.~\ref{fig:decomposition_mass_evolution_normalised}, there is some degree of morphological diversity, which can be tracked to differences in the growth and evolution of the stellar mass and particular features such as disc instabilities and merger events (see also \citealt{Iza2022,Iza2024}).

\begin{figure}
    \centering
    \includegraphics[draft=false]{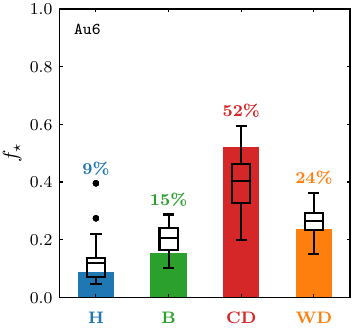}
    \caption{
    Fraction of the total stellar mass of the galaxy within each galactic component for Au6 at $z=0$ (coloured bars).
    The box plots indicate the statistical properties of the full sample of galaxies.
    Each box extends from the first quartile to the third quartile and the line inside indicates the value of the median.
    The whiskers show the values of the farthest data point that is within $3/2$ of the inter-quartile range (the size of the box) from the box.
    Black data points are the values that extend beyond the end of the whiskers.
    }
    \label{fig:stellar_mass_distribution_au6}
\end{figure}

At $z=0$, most of the stellar mass in Au6 belongs to the disc components (76\%, with 52\% in the cold disc and 24\% in the warm disc) while the spheroidal components concentrate only 24\% of all stellar mass.
In Fig.~\ref{fig:stellar_mass_distribution_au6} we compare Au6 with the rest of the galaxies.
The statistical properties included in the figure (see caption) show that there is a general trend for the Auriga galaxies in terms of the relative amount of stellar mass in the different components.
The most significant variations are found for the stellar haloes and the discs, which are more strongly affected by mergers, gas accretion and instabilities.
It is worth noting that Au6, our reference galaxy, exhibits stellar fractions in the bulge and cold disc outside the two central quartiles of the distributions.

\begin{figure}
    \centering
    \includegraphics[draft=false]{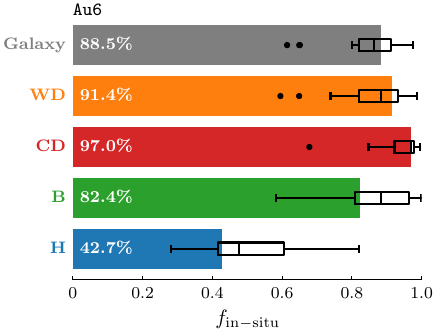}
    \caption{
    Fraction of the stars in each component of Au6 at $z=0$ that were born in the galaxy (in situ stellar fraction).
    For reference, we also show the box plot as in Fig.~\ref{fig:stellar_mass_distribution_au6}.
    }           
    \label{fig:stellar_origin_by_component_au6}
\end{figure}

Another important descriptor for the stellar populations which can affect the resulting chemical properties of galaxies is their origin.
For this reason, we investigate separately the in situ (stars formed in the main progenitor; i.e., stars that are in the main object when they are first detected in the simulation) and ex situ (stars formed in systems other than the main progenitor; i.e., stars that are in an object other than the main one when they are first detected in the simulation) stellar populations.
Fig.~\ref{fig:stellar_origin_by_component_au6} shows the fraction of in situ stars in Au6, both for the whole galaxy and separately for each stellar component.
We also include statistical measurements of the full sample.
As is expected for a galaxy that has not suffered important mergers in the latest epochs of its evolution, almost $90\%$ of all stars in Au6 formed in situ, and this fraction goes up to $97\%$ and $91.4\%$, respectively, for the cold and warm discs.
The in situ fraction of the bulge is still relatively high ($82.4\%$), while the halo has only $42.7\%$ of in situ stars.
The rest of the galaxies show a similar behaviour, with the general tendency of having high in situ fractions for the cold and warm discs\footnote{See \cite{Gomez2017} for a study on the ex situ discs in the Auriga simulations.}, lower but similar fractions for the bulges (consistent with the results obtained by \citealt{Gargiulo2019} for the bulges of the same set of simulations), and with the haloes having a larger contribution from ex situ stars (consistent with the results obtained by \citealt{GonzalezJara2025}, and \citealt{Monachesi2019} for the halo).
The most important variations are found for the stellar haloes, which present from relatively low ($\sim 30\%$) to quite high ($\sim 80\%$) ex situ.

It is worth mentioning that there are few galaxies in which the in situ fractions are higher for the bulges compared to the cold discs (Au8, Au14, Au17, Au20, and Au24).
Interestingly, these systems are either currently being subject to merger events (such as Au8, Au20, and Au24), have discs that have been (partially) destroyed in the past (such as Au14) or have developed strong galactic bars (such as Au17) \citep{Iza2024}.
Merger events and the development of galactic bars\footnote{See \cite{Fragkoudi2025} for a study on the formation and evolution of galactic bars in the Auriga simulations.} can have considerable influence on the evolution of galaxies.
For instance, galactic discs with large ex situ stellar populations are expected to form as a result of massive mergers with low angles of incidence (see \citealt{Gomez2017} for further details on the angles of incidence in the Auriga galaxies and their impact on disc development).

\begin{figure*}
    \centering
    \includegraphics[draft=false]{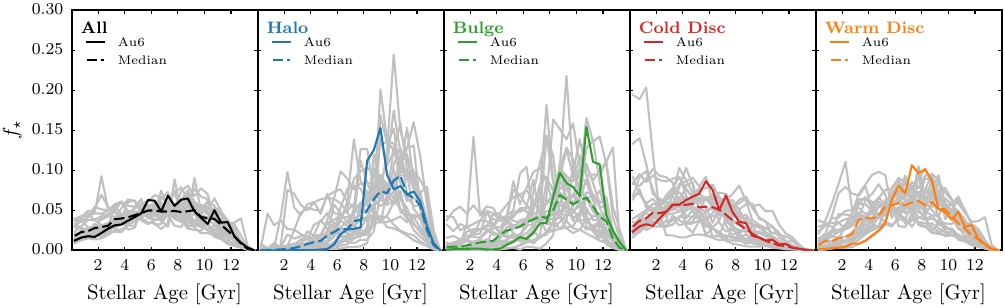}
    \caption{
    Fraction of the stars with a given stellar age in Gyr at $z=0$.
    Each panel shows the distribution for a given galactic component, normalised to the amount of the stars in that component.
    The panel on the left shows the distribution of all stars in the galaxy, normalised to the total number of stars.
    In each panel, we highlight Au6 and the median of the sample.
    }
    \label{fig:stellar_age_distribution_au6}
\end{figure*}

Figure~\ref{fig:stellar_age_distribution_au6} shows the distribution of stellar ages for Au6, for all stars and separated into the different components (coloured lines), overlaid to the rest of the sample (grey lines).
The bulge and stellar halo components are the two oldest populations, opposite to the cold disc which is formed by younger stars.
Most stars in the bulge and stellar halo of Au6 have ages in the range $6$--$14~\mathrm{Gyr}$, the warm disc stars have ages of range $4$--$12~\mathrm{Gyr}$, and the cold disc contains the youngest population with ages in the range $0$--$9~\mathrm{Gyr}$.
We find similar trends for all galaxies -- the average relation is similar to the results for Au6 -- with the largest variations detected for the cold discs, which in a few cases exhibit a significant contribution of very young stars and, to a lesser extent, in the bulges which also, in some cases, have significant amounts of young stars in addition to the old component.

In order to quantify the spatial extent of the components in each galaxy, we calculated the $90^\mathrm{th}$ mass-weighted percentiles of the spherical radius for the bulge and halo ($R^\mathrm{b}$ and $R^\mathrm{h}$, respectively) and, for the discs, the corresponding $90^\mathrm{th}$ percentiles of the absolute value of the height above the disc plane ($h^\mathrm{cd}$ and $h^\mathrm{wd}$), and of the cylindrical radius ($r_{xy}^\mathrm{cd}$ and $r_{xy}^\mathrm{wd}$). 
At $z=0$, the values for the median over the full sample are $\tilde{R}^\mathrm{b} = 4.4~\mathrm{kpc}$,
$\tilde{R}^\mathrm{h} = 68.7~\mathrm{kpc}$ for the bulge and halo, $\tilde{h}^\mathrm{cd}= 2.6~\mathrm{kpc}$ and $\tilde{r}_{xy}^\mathrm{cd} = 17.7~\mathrm{kpc}$ for the cold disc, and $\tilde{h}^\mathrm{wd} = 5.2~\mathrm{kpc}$ and $\tilde{r}_{xy}^\mathrm{cd} = 19.5~\mathrm{kpc}$ for the warm disc. 
We note that the warm disc is thicker than the cold disc by a median factor of $\sim 2$, and the two discs do not differ significantly in radius.

\section{The distribution of metals today}
\label{sec:metals}

Once the galactic decomposition has been defined and properly tested, we investigate the chemical composition of the various components at $z=0$, focusing on the distributions of [Fe/H], [O/H] and [O/Fe], the age-metallicity relation, and the metallicity profiles in the cold discs.
As in the previous section, we use Au6 as our example galaxy and compare results to the full Auriga sample.

\subsection{The age-metallicity relation}

\begin{figure*}
    \centering
    \includegraphics[draft=false]{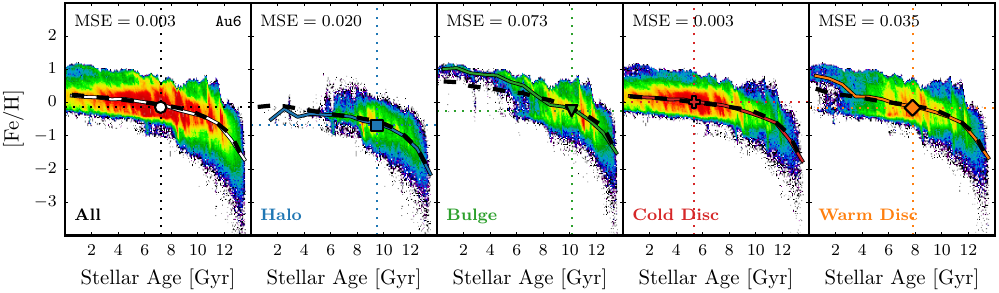}
    \caption{
    Age-metallicity relation for the stars of Au6 at $z=0$.
    In the $x$-axis we indicate the stellar age in Gyr and in the $y$-axis we show the [Fe/H] abundance ratio.
    The colour map indicates the amount of stars in each pixel in a logarithmic scale.
    Each panel shows the distribution of the stars that belong to a specific galactic component: all galaxy stars (first panel), halo (second panel), bulge (third panel), cold disc (fourth panel), and warm disc (fifth panel).
    In each panel, the symbol indicates the location of the median.
    Furthermore, we also indicate the metallicity trend as a function of stellar age for all galaxies in the sample (black dashed line) and the median for Au6 (continuous line, colour-coded according to each component); on the top left, we show the mean squared error between these two lines.
    }
    \label{fig:age_metallicity_by_region_au6_or_l4_s127}
\end{figure*}

Figure~\ref{fig:age_metallicity_by_region_au6_or_l4_s127} shows the age-metallicity relation, at $z=0$, for all stars in Au6 (on the left panel), and separately for each galactic component (subsequent panels).
Au6 stars present a well-defined correlation, showing how newer generations of stars acquire increasingly higher metallicities.
As expected, old stars have a wider range of [Fe/H] values of $\sim [-3, 0]$ and, as we move towards the younger population of the disc, stars are more metal-rich and the distribution becomes narrower.
We find the differences in the enrichment histories of the different components, with the stellar halo and bulge having a rapid increase in metallicity at early times, and the discs experiencing a smoother, continuous enrichment which encompasses the formation of these components over longer timescales.

From this figure we can also observe that the mean stellar ages of the bulge and halo components (vertical dotted lines) are of the order of $10~\mathrm{Gyr}$, while those of the cold and warm discs are $\sim 6~\mathrm{Gyr}$ and $\sim 8~\mathrm{Gyr}$, respectively.
These trends in mean stellar ages reflect in the different (mean) levels of [Fe/H] (horizontal dotted lines): the stellar halo is the least enriched component, with a mean [Fe/H] of $\sim -0.6$, the bulge has an intermediate metallicity of $\mathrm{[Fe/H]} \sim -0.1$, and the younger components in the discs have near solar (median) metallicities.

In Fig.~\ref{fig:age_metallicity_by_region_au6_or_l4_s127} we also include the mean age-metallicity relation for Au6 (coloured lines), together with that of the full Auriga sample (dashed black lines).
From these lines, we can see that Au6 lies near the mean value over the full sample, although it presents slightly higher [Fe/H] values in the bulge for the young population.
We also include in the figure the mean squared error between the mean age-metallicity relations of Au6 and the full sample, finding the lowest value for the cold disc and the highest one for the bulge.

\begin{figure*}
    \centering
    \includegraphics{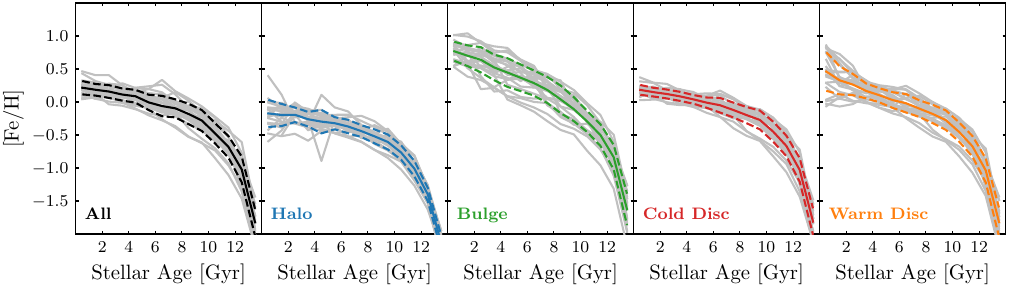}
    \caption{
    Age-metallicity relation for the Auriga sample.
    In the $y$-axis we show the median [Fe/H] metal abundance in each stellar age bin and the panels represent the stars in different components (the panel on the left represents all stars).
    Grey lines represent the curves for all galaxies in the sample, while the solid coloured line shows the curve for the sample average.
    The dashed line in each panel shows the $\pm1$ standard deviation of the sample average.
    }
    \label{fig:amr_sample_median}
\end{figure*}

Figure~\ref{fig:amr_sample_median} analyses the homogeneity of the Auriga sample in terms of the age-metalliticty relation.
This figure shows the $z=0$ age-metallicity relation for all galaxies (grey lines), considering all stars and also separated by stellar component.
The mean relation is shown as a solid line and the dashed lines indicate the $\pm 1 \sigma$ levels, being $\sigma$ the standard deviation.
From this figure, it is clear that the cold disc is the most homogeneous component (with a profile-averaged standard deviation of 0.114 versus 0.122, 0.188 and 0.161 for the halo, bulge and warm disc, respectively), with small galaxy-to-galaxy variations.
In contrast, the largest differences are found for the bulges.
As we show below, these differences translate into variations in the stellar abundance distributions.
Finally, it is worth noting that we do not observe evidence of high perturbations in the age-metallicity relation for any galaxy in the sample, even for those with the most prominent merger events.
The similar levels of enrichment for all Au galaxies are also consistent with a mass-metallicity relation at the MW-mass scale, and with the relatively quiet merger history of galaxies in the sample.

\subsection{Abundance ratio distributions}

\begin{figure*}
    \centering
    \includegraphics[draft=false]{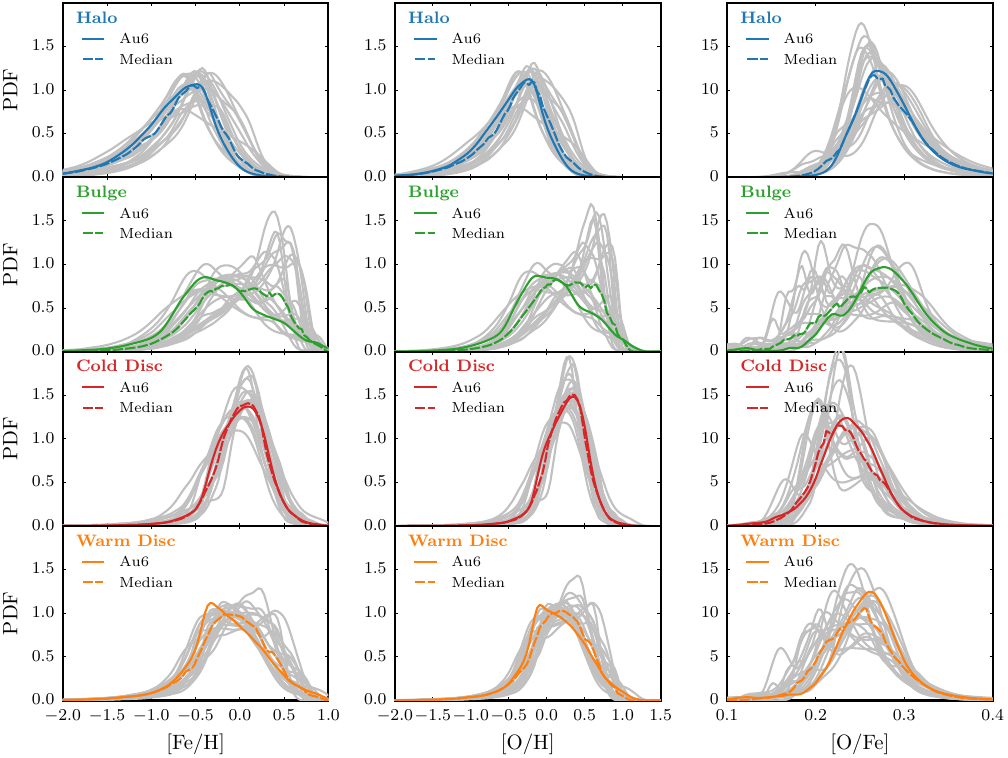}
    \caption{
    Distribution of the [Fe/H] (left), [O/H] (middle), and [O/Fe] (right) metal abundances for the Auriga sample.
    Each row shows a different galactic component, as indicated in the top left.
    In each panel, grey curves show the distribution of each galaxy in the sample, highlighting Au6 in colour.
    }
    \label{fig:abundance_dist}
\end{figure*}

\begin{figure*}
    \centering
    \includegraphics{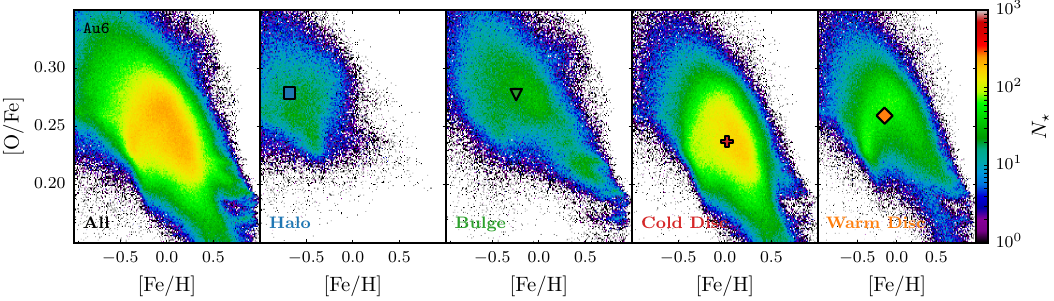}
    \caption{
    Distribution of stars in the [O/Fe]-[Fe/H] plane for Au6 at $z=0$.
    The colour map indicates the amount of stars per pixel, in a logarithmic scale spanning three orders of magnitude.
    Each panel shows the distribution for a different component (with the first panel showing the distribution for all stars).
    The symbols indicate the location of the median for each galactic component, as indicated at the bottom left corner of each panel.
    }
    \label{fig:map_OFe_vs_FeH_Au6}
\end{figure*}

\begin{figure}
    \centering
    \includegraphics{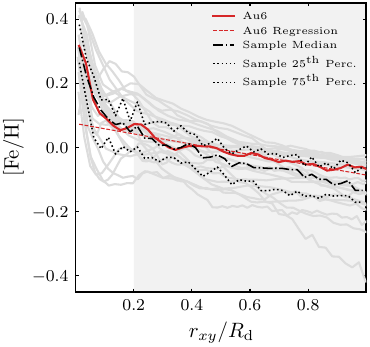}
    \caption{
    [Fe/H] abundance profiles for the stars in the cold disc.
    Each grey line represents a galaxy in the sample, with Au6 highlighted in red and its linear regression shown as a dashed line.
    We also show the median over the sample and the 25$^\mathrm{th}$ and 75$^\mathrm{th}$ percentiles.
    The linear regression is performed in the region $0.2 R_\mathrm{d} \leq r_{xy} \leq R_\mathrm{d}$ (shown as a grey area for reference).
    }
    \label{fig:feh_abundance_profiles_sample}
\end{figure}

\begin{figure*}
    \centering
    \includegraphics{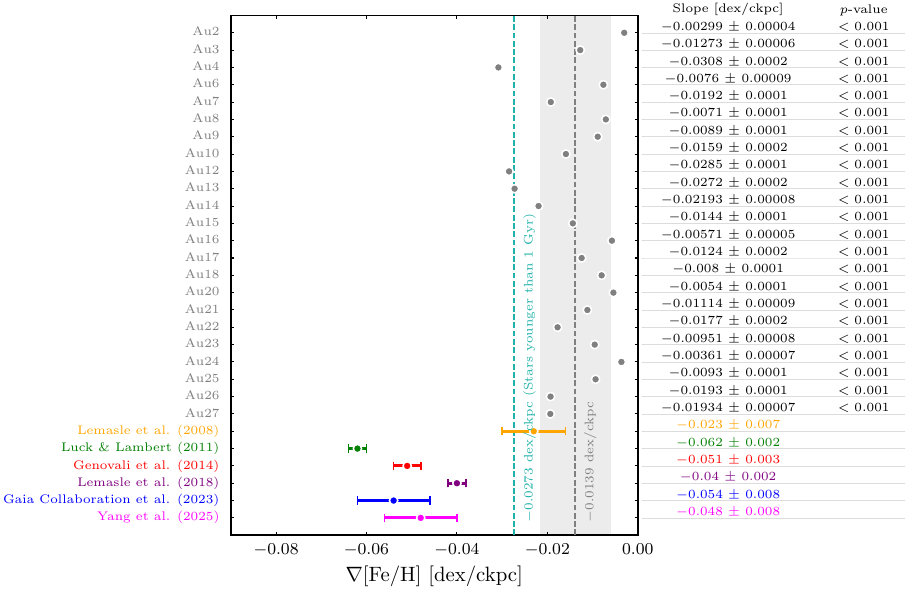}
    \caption{
    Comparison of the [Fe/H] abundance gradients for the stars in the cold disc, corresponding to the linear regressions of Fig.~\ref{fig:feh_abundance_profiles_sample}, to available observational estimates.
    The vertical dashed line indicates the average value of the Auriga sample, for all stars (grey) and stars younger than $1~\mathrm{Gyr}$ (cyan).
    On the right, we indicate the value of each gradient and its standard error along with the $p$-value of the fit.
    }
    \label{fig:feh_abundance_gradients}
\end{figure*}

The left and middle panels of Fig.~\ref{fig:abundance_dist} show the $z=0$ distribution of the [Fe/H] and [O/H] abundances for galaxy Au6 as probability density functions (PDF), for the different stellar components.
We use the oxygen abundance as a proxy for the behaviour of the $\alpha$-elements, which all show very similar distributions despite variations in the overall levels.
We also include results for the rest of the Auriga sample, as grey lines, in order to study the homogeneity of the sample.

We find that, in the case of Au6, all components exhibit single-peak, bell-shaped distributions (in the case of the bulge, we also find a second, under-dominant peak).
Consistent with our previous results, the peak of the distribution appears at a lowest (highest) metallicity for the stellar halo (cold disc), both for iron and oxygen.
Similar trends are found if we look at the full sample (grey lines), although significant galaxy-to-galaxy variations are detected for the bulges, where a double-peaked distribution is present in many cases.
The high-metallicity peaks are, in fact, the result of significant recent star formation in the bulge of these galaxies, which acquire higher metallicities compared to the old bulge population (see Fig.~\ref{fig:stellar_age_distribution_au6}).
While the bulges are, in fact, the less homogeneous component, we also observed that the warm discs for the full sample show some variation.
However, this could also be due to potential contamination from bulge particles, as the typical abundances agree with those of the bulges.

\subsection{The [O/Fe] versus [Fe/H] relation}

In the right-hand panel of Fig.~\ref{fig:abundance_dist} we show the corresponding distributions for the abundance ratio [O/Fe].
In this case, we observe how the oldest components in Au6 -- stellar halo and bulge -- are the most $\alpha$-enhanced, as a result of the rapid contamination with $\alpha$-elements originated in SNII events which occurred early on, where iron has not yet been formed in significant amounts.
The warm disc is slightly less, but similarly enriched, compared to the bulge and halo, and the cold disc, being the youngest component, peaks at lower [O/Fe].
Similar trends are found for the rest of the galaxies; following the differences found in [Fe/H] and [O/H], with the bulges presenting the less homogeneous, more complex [O/Fe] distribution.

In Fig.~\ref{fig:map_OFe_vs_FeH_Au6} we show the distribution of stellar metallicities in the [O/Fe]--[Fe/H] plane for Au6, at $z=0$.
We also include symbols at the median values for the stars in each component.
Each pixel is coloured according to the star count, using a logarithmic scale that spans three orders of magnitude.
Stellar halo stars locate in the low [Fe/H], high [O/Fe] regime, due to their old ages, in contrast to the young stars in the disc which are more enriched -- with near-solar metallicities -- and less $\alpha$-enhanced.
The bulge has a similar median [O/Fe] value compared to the stellar halo, and a higher [Fe/H] ratio, and the warm disc's median abundance ratios locate in between the cold disc and the bulge.
The distribution of stars in the [O/Fe]--[Fe/H] plane found for Au6 is similar to that observed in the other Auriga galaxies, as we discuss in more detail in Section~\ref{sec:global_trends}.

The behaviour observed in the [O/Fe]--[Fe/H] plane is the result of the different processes involved in the production of metals in the simulations and their ejection at different time-scales.
Most metals are produced and injected into the interstellar medium by massive, short-lived stars ($\sim 10^{7} ~\mathrm{yr}$) which end their lives as SNII, except for iron which is primarily produced by SNIa and injected into the medium over longer timescales of $\sim 10^{8}$--$10^{9}~\mathrm{yr}$ \citep{Mosconi2001, Tornatore2004, Scannapieco2005}.
Older populations are thus expected to be $\alpha$-enriched, as these are contaminated with metals prior to iron synthesis. In contrast, younger stars form from an evolved medium already contaminated by iron and thus typically have lower [O/Fe] ratios.

\subsection{[Fe/H] metallicity profiles}

As we discussed in the previous sections, the cold disc is the most homogeneous component in terms of the metal content when we look at the different Auriga galaxies, and -- despite minor differences in the mean stellar age of each system -- the most dominated by young stars.
The homogeneity in the chemical history is in part due to the selection of the Auriga galaxies, which have quiet merger histories, particularly in the recent past.
However, we have shown in previous work \citep{Iza2022, Iza2024} that there are important differences in the growth of discs among the various galaxies as a result of particular features in the formation history of each system.
For example, some discs formed very early on and stayed stable up to $z=0$, while other systems experienced episodes of partial/total disc destruction, in some cases forming new discs later on.
We have also shown in \cite{Iza2024} that most galaxies formed their discs from the inside-out, and that this resulted from an inside-out pattern in the accretion history.
In this section, we investigate the radial metallicity profiles of the discs in the Auriga galaxies.

Fig.~\ref{fig:feh_abundance_profiles_sample} presents the $z=0$ stellar metallicity profile through the [Fe/H] abundance ratio for all galaxies in the sample (grey lines) and for Au6 (red line), using stars identified as part of the cold discs.
We also include the median relation and some statistical measurements of the sample.
We can see that Au6 is not very far from the median of the sample.
In the figure, we also display a linear fit to the Au6 relation, that has been performed between $0.2$ to $1$ times the disc radius, $R_\mathrm{d}$, calculated as that enclosing $90\%$ of the stellar mass (see \citealt{Iza2022}).
We use these limits in the dynamic decomposition to avoid possible contamination of spheroidal stars belonging to the bulge and the stellar halo (typically located in the inner and outer regions of the disc).
Most galaxies have central abundance values in the range [Fe/H] $\sim 0.2$--$0.4~\mathrm{dex}$, that decrease rapidly in the inner regions, and define an approximately linear, smooth decay up to the disc radius where all galaxies have sub-Solar abundances, between $-0.2$ and $-0.4~\mathrm{dex}$.
Linear fits to each galaxy (starting at $0.2\, R_\mathrm{d}$) produce slopes in the range $\nabla {\mathrm{[Fe/H]}}\sim[ -0.03,-0.003] ~ \mathrm{dex\, kpc^{-1}}$, with Au6 having a slope of $-0.0076\pm 0.0001~\mathrm{dex\, kpc^{-1}}$, which is not so far from the mean relation of the sample that gives $-0.0139~\mathrm{dex\, kpc^{-1}}$.

Different observational measurements of the slope of the [Fe/H] profile in the MW have been obtained, for instance: $-0.023\pm 0.007~\mathrm{dex\, kpc^{-1}}$ \citep{Lemasle2008}, $-0.062 \pm 0.002~\mathrm{dex\, kpc^{-1}}$ \citep{Luck2011}, $-0.051\pm 0.003~\mathrm{dex\, kpc^{-1}}$ \citep{Genovali2014} and $-0.04\pm 0.002~\mathrm{dex\, kpc^{-1}}$ \citep{Lemasle2018} using galactic Cepheids, and $-0.054 \pm 0.008~\mathrm{dex\, kpc^{-1}}$ \citep{Gaia2023} and $-0.048 \pm 0.008~\mathrm{dex\, kpc^{-1}}$ \citep{Yang2025} using open clusters.
It is worth noting that the fact that different measurements do not agree with each other reflects the complexity in obtaining these data, as well as the biases and systematic effects affecting observational estimates.
We find that, except for \cite{Lemasle2008}, the slopes obtained for the simulated galaxies are inconsistent with observations, which in general show steeper profiles.
However, a subset of $13$ of the Auriga galaxies (Au7, Au10, Au12, Au13, Au14, Au15, Au17, Au18, Au21, Au22, Au25, Au26, and Au27) agrees well with the result of \cite{Lemasle2008} at a $1 \sigma$ level.
This can be better seen in Fig.~\ref{fig:feh_abundance_gradients} where a detailed comparison of the individual cold disc metallicity gradients of all Auriga galaxies to observational estimates is shown.

Of the 13 galaxies whose profiles are consistent with \cite{Lemasle2008}, all but three (Au7, Au12, and Au15) were previously identified as MW-like systems when studying their formation and merging history \citep{Iza2022}, and all but five (Au10, Au13, Au14, Au17, Au22) were also consistent with the inside-out formation scenario \citep{Iza2024}.
This indicates that the galaxies with gradients most similar to the one found by \cite{Lemasle2008} are those with evolution histories that are closer to the standard formation scenario for the MW, which consists of a quiescent merger history in the last several billion years that helped the development of the thin disc.

In Fig.~\ref{fig:feh_abundance_gradients} we also indicate the average metallicity gradient  obtained when only young stars (ages below $1~\mathrm{Gyr}$) are considered.
In this case, the slope is more pronounced, with a value of $-0.0273~\mathrm{dex\, kpc^{-1}}$, moving towards the observational measurements.
This change in the average slope is due to the increase in the mean metallicity near the centre of the galaxy, when old stars are not considered.

\section{Global trends in the chemical patterns of the different components}
\label{sec:global_trends}

\begin{figure*}
    \centering
    \includegraphics{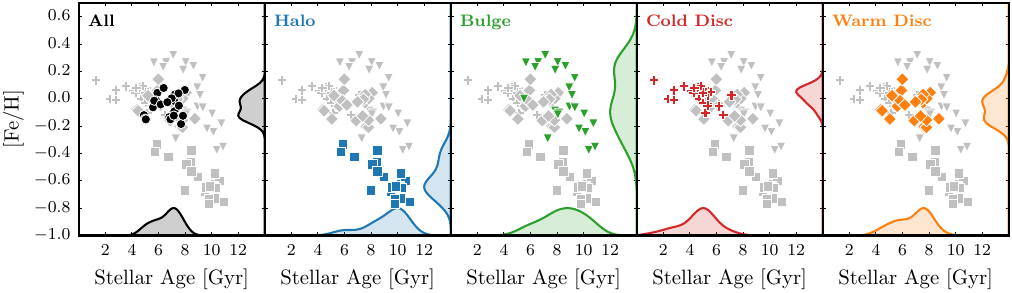}
    \includegraphics{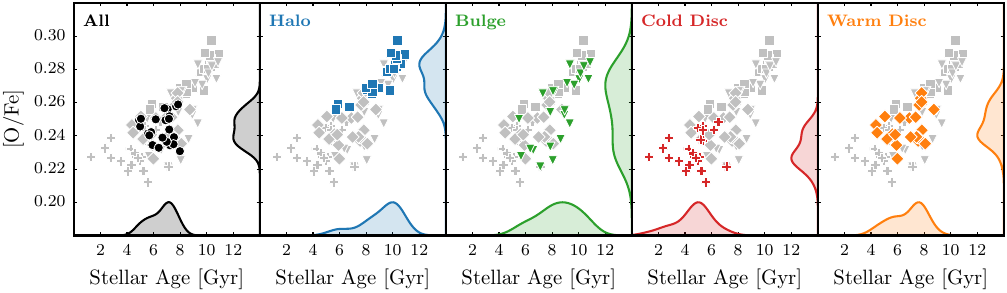}
    \includegraphics{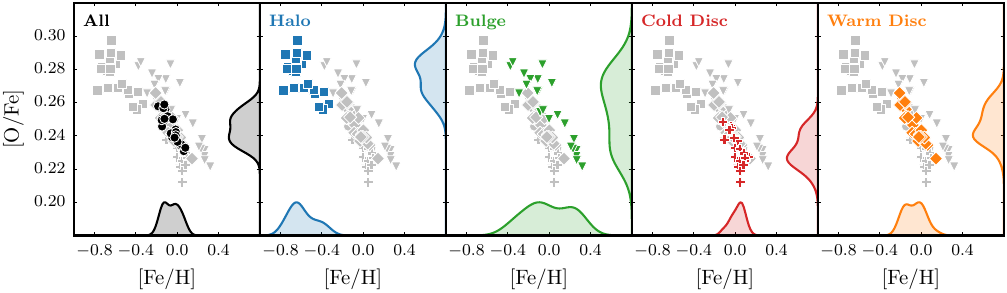}
    \caption{
    Top: Median [Fe/H] abundance vs. median stellar age at $z=0$.
    Each point in these scatter plots is a galaxy, and each panel highlights one galactic component in colour.
    Halo, bulge, cold disc, and warm disc are shown in blue, green, red and orange, respectively.
    The grey symbols in the background show the values for all components.
    For reference, we also show the probability distribution for each axis as a filled line plot.
    Middle: Same as the top panels but for the [O/Fe] abundance.
    Bottom: Median [O/Fe] abundance vs. median [Fe/H] abundance at $z=0$.
    Each point in these scatter plots is a galaxy, and each panel highlights one galactic component in colour.
    Halo, bulge, cold disc, and warm disc are shown in blue, green, red and orange, respectively.
    The grey dots in the background show the values for all components.
    For reference, we also show the probability distribution for each axis as a filled line plot.
    }
    \label{fig:sample_correlations}
\end{figure*}

In the previous section, we analysed the chemical patterns of the stars in the four stellar components identified in Au6, which has properties close to the median over the Auriga galaxy sample.
Now, we turn our focus to the global trends of each component.
For this, we compute the median values of stellar age, [Fe/H] and [O/Fe] for each simulation and stellar component, and investigate similarities and differences between the model galaxies.

Figure~\ref{fig:sample_correlations} shows various correlations between the mean [Fe/H] ratios, the [O/Fe] ratios, and the mean stellar ages for the Auriga galaxies at $z=0$.
We also include the corresponding probability distributions in the $x$- and $y$-axes.
We show results considering all stars in the galaxies, as well as separated by stellar component.
When we consider all stars in each galaxy, we find that all systems have similar values for the mean [Fe/H] and [O/Fe] ratios, with less than $0.5~\mathrm{dex}$ galaxy-to-galaxy variations.
We also find similar mean ages in the range $5$--$8~ \mathrm{Gyr}$.
We find no correlation between the mean abundance ratios and the mean ages, but there is a clear anti-correlation between [Fe/H] and [O/Fe] ratios, as expected.
This results from the combined effects of the slow ejection of iron into the ISM by SNIa and the rapid contamination by $\alpha$-elements like oxygen as a result of SNII \citep{Tinsley1979}.

When looking at each stellar component separately, we observed that they are systematically located at different regions in the various relations, indicating general trends that are valid for all galaxies in the sample.
The stellar haloes are always the least enriched, are more $\alpha$-enriched, and are the oldest component.
The bulges have higher metallicities, a wider range of [O/Fe], and intermediate stellar ages.
The warm and cold discs are young, metal-rich, and have low $\alpha$-enrichment levels.
The warm and cold discs are also the most homogeneous components, and the bulges exhibit the largest scatter, particularly in the abundance ratios.
The anti-correlation in the [Fe/H]-[O/Fe] relation observed for all stars appears even more clearly when we look at the different components separately.

\begin{figure*}
    \centering
    \includegraphics[draft=false, scale=0.75]{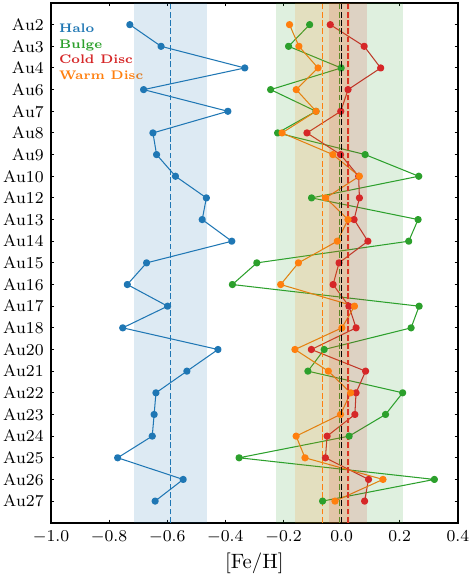}
    \includegraphics[draft=false, scale=0.75]{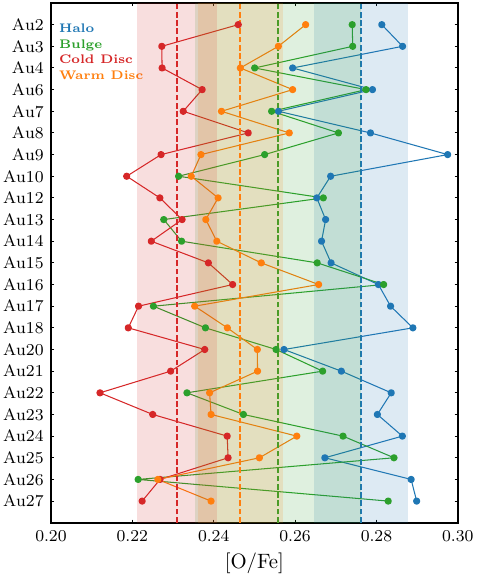}
    \includegraphics[draft=false, scale=0.75]{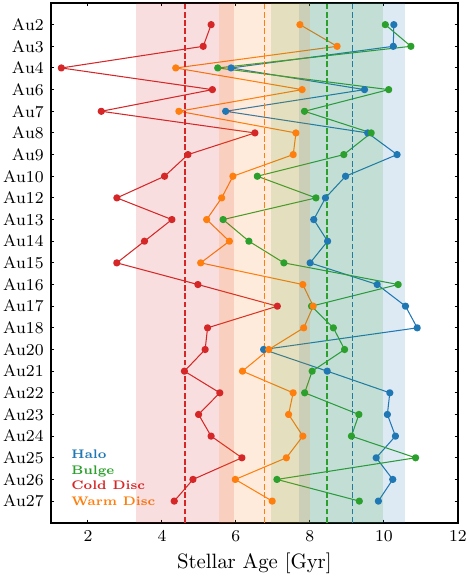}
    \caption{
    Mean values of [Fe/H], [O/Fe] and stellar age of all the galaxies in the sample, ordered by galaxy.
    Each dot indicates the value of an observable for a given galaxy, colour-coded by component, while the dashed vertical lines shows the average over the sample.
    Shaded areas indicate the $\pm 1 \sigma$ region.
    We note that the solid lines that join the data points are included for visualisation purposes but lack any intrinsic meaning.
    }
    \label{fig:sample_ordered_iron_abundance}
\end{figure*}

In order to better understand if there are systematic relations between the abundances of the different components in each individual galaxy, we show in Fig.~\ref{fig:sample_ordered_iron_abundance} the median [Fe/H], [O/Fe] and stellar age values for all simulations.
In each case, we also include the corresponding averages over the whole sample (vertical dashed lines) and shaded regions indicating the $\pm 1 \sigma$ levels.
In each galaxy, the stellar halo is, with very few exceptions, the least enriched, most $\alpha$-enhanced, and oldest component.
The warm disc of each galaxy also appears systematically at lower [Fe/H] values and higher [O/Fe] values compared to the cold disc, and at larger stellar ages. 
Finally, and consistently with our previous findings, the bulge is systematically older than the cold disc, and generally more $\alpha$-enhanced, but significant galaxy-to-galaxy variations are detected when we compare the [Fe/H] abundances of the bulge and cold disc of each galaxy.

We found no correlation between the abundances of the different components and other galactic properties such as stellar and virial masses, bulge-to-total and cold disc-to-total mass ratios, bulge, cold disc and stellar ages, and [Fe/H] gradient.
Although our sample shows no significant galaxy-to-galaxy variations in any of these properties, we have shown in previous work \citep{Iza2022, Iza2024} that there are differences in the growth and evolution of the discs, and in the amount and time of merger episodes.
Our results therefore indicate that, despite these differences, MW-like galaxies are expected to have similar iron abundances and $\alpha$-enrichment levels, and that there are systematics in the relative abundances of the different stellar components, except in the case of the bulges that are quite diverse.

\section{The origin of chemical patterns}
\label{sec:origin}

In previous sections, we described the chemical properties in the galaxies of the Auriga project at $z=0$.
We studied typical distributions of stellar abundances, ages, abundance ratios and metallicity gradients in different galactic components identified using a simple dynamical decomposition based on the circularity parameter and the normalised potential.
We also studied galaxy-to-galaxy variations in the chemical abundances as well as systematic relations in the abundances of the different stellar components.
In this section, we want to investigate the origin of the chemical patterns previously found.
For this, we performed our galactic decomposition for all galaxies and at all times, investigating stellar migration -- from one component to another -- between the stellar birth time and $z=0$.
As the chemical properties of stars are mainly determined at the birth time, if exchange of stars between components is important, it could leave imprints on the $z=0$ chemical properties.
It is worth noting that this analysis can only be done for in situ stars, which can be assigned to one of the stellar components of the main progenitor at the time of birth.
In order to better understand the general trends of mass exchange between the different stellar components, here we show average results for the whole Auriga sample.

\begin{figure*}
    \centering
    \includegraphics{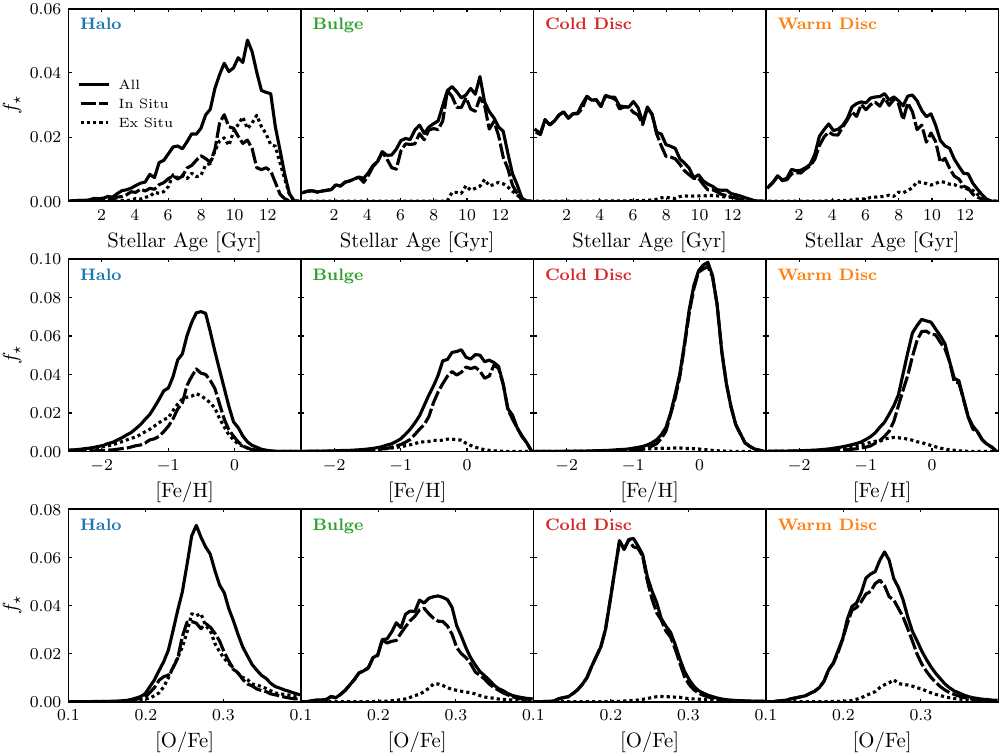}
    \caption{
    Median distribution of stellar ages, [Fe/H] abundance, and [O/Fe] abundance for the sample of galaxies.
    Top: Fraction of stars of a given age for all stars in the galaxy and stars identified as in situ and ex situ, as indicated in the legend.
    Each panel shows the distribution that corresponds to a given component (halo, bulge, cold disc, and warm disc), normalised to the total amount of stars in that component.
    Middle: Same as the top row, but for the [Fe/H] abundance.
    Bottom: Same as the top row, but for the [O/Fe] abundance.
    }
    \label{fig:present_day_distributions}
\end{figure*}

We first investigated the level of influence of ex situ stars, in order to confirm that excluding them from our migration analysis has no impact on our results.
In Fig.~\ref{fig:present_day_distributions}, we show the median distributions of stellar ages and [Fe/H] and [O/Fe] abundance ratios, at $z=0$, together with the distributions of the in situ and ex situ components separately.
The ex situ stars are typically old -- the vast majority being older than $\sim 8~\mathrm{Gyr}$ -- and have sub-Solar [Fe/H] values and [O/Fe] abundances in the range $\sim 0.25$--$0.35$.
They comprise, on average, a small fraction of the stellar mass for the bulge and for the cold and warm discs, and therefore, they are not expected to have any significant impact on the global chemical properties of these components\footnote{Even though ex situ stars do not have a significant impact on the global chemical properties of these galaxies, they are typically located at a distinctive region in the age-metallicity plane \citep{Gomez2017}.}.
In contrast, around $50\%$ of the stellar halo, on average, is composed of ex situ stars.
In this case, the stellar age, [Fe/H] and [O/Fe] distributions are similar for the in situ and ex situ stars, and thus it is also expected that ignoring them in our migration analysis will have no significant effect on our conclusions.
It is anyway worth mentioning that any signature of ex situ stars on the chemical properties of galactic systems should be expected to show more clearly in the metal-poor population.

\begin{figure*}
    \centering
    \includegraphics{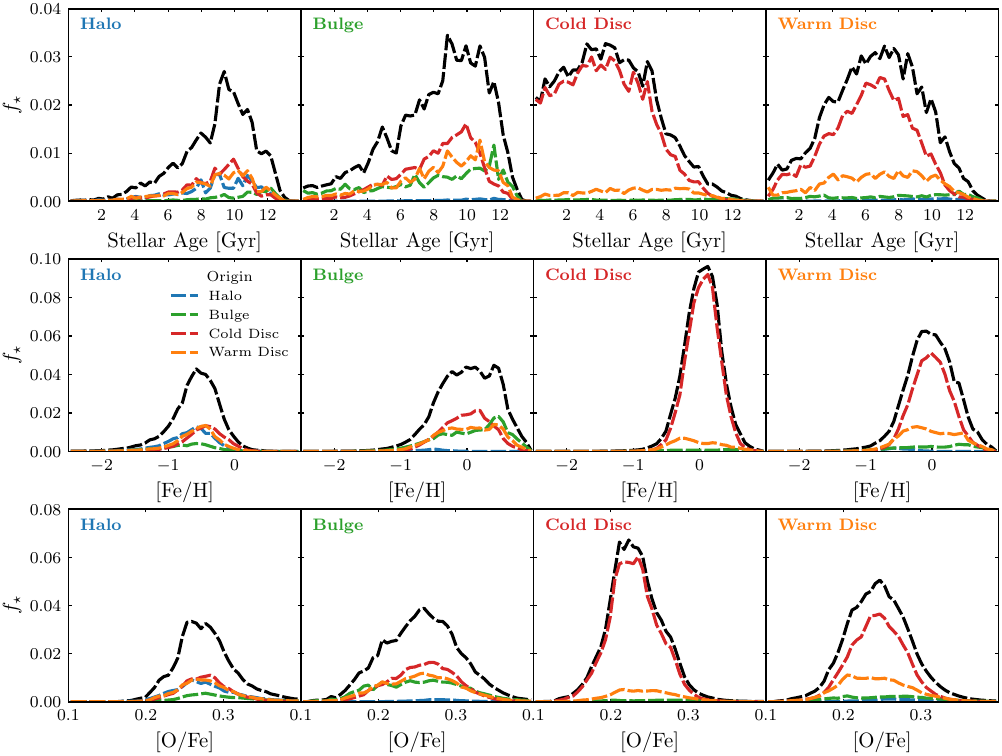}
    \caption{
    Median distribution of stellar ages, [Fe/H] abundance, and [O/Fe] abundance for the sample of galaxies, separated by the component at birth.
    In each panel, the black dashed line shows the distribution of the in situ stars for each component for reference.
    This line corresponds to the line with the same style in Fig.~\ref{fig:present_day_distributions}.
    Top: Distribution of the stars that reside in each component at $z=0$ (in situ stars), separated by the component they inhabited at the time of birth.
    The red line in the first panel, for example, shows the distribution of stellar ages for the stars that populate the halo at $z=0$ but that were born in the cold disc.
    The green line in the fourth panel, on the other hand, shows the distribution of stellar ages for the stars that reside in the warm disc at $z=0$ but were born in the bulge.
    Middle: Same as the top row, but for the [Fe/H] abundance.
    Bottom: Same as the top row, but for the [O/Fe] abundance.
    }
    \label{fig:origin_distributions}
\end{figure*}

We now analyse possible stellar migration between the time of birth and $z=0$ for the in situ population.
Fig.~\ref{fig:origin_distributions} shows the median $z=0$ distributions of stellar age, [Fe/H] and [O/Fe] for the in situ stars, together with the relative contribution of stars born in the different stellar components (i.e. the component of each star when the particle is first detected in the simulation, with the galactic decomposition computed at that time)\footnote{Although we focus on the median for this analysis, the distributions shown in Fig.~\ref{fig:origin_distributions} are also representative of the individual galaxies in the Auriga sample.}.
In the case of the stellar halo, we find similar contributions from stars that were born in the cold disc, the warm disc and the halo, with similar distributions of [Fe/H] and [O/Fe].
As expected, there is very little contribution to the present-day stellar halo from stars born in the bulge, which in any case have similar abundances than the rest of the stellar halo population.

A similar trend is found for the bulge, with stars having been born either in it or in the cold and warm discs, with similar proportions.
In this case, there is a negligible contribution from stars born in the stellar halo.
Stars born in the bulge populate, as expected, the high [Fe/H] and low [O/Fe] tails of the abundance distributions, and stars born in the cold and warm discs contribute with similar abundances.
In general terms, the three components of origin -- cold disc, warm disc and bulge -- contribute with similar chemical distributions to the $z=0$ bulge stars.
Our results are in agreement with \cite{Boin2024} who found, by analysing both observational and simulated data, that metal-rich and metal-poor stars in the bulge of the MW mainly originate from the thin and thick discs, and with those of \cite{Gargiulo2019} who found that the bulges of the Auriga simulations are mainly composed of in situ stars.

A different behaviour is found for the warm and cold discs, where we find a negligible contribution from stars born in the stellar halo and a small contribution from stars born in the bulge.
Moreover, while most stars that are in the present-day cold disc were also in this component at birth, a significant fraction of the stars that populate the warm disc at $z=0$ were in the cold disc component at their birth-times.
This implies that the chemical history of material near the disc plane might have also influenced the resulting abundances of the warm disc (which, as shown above, is always thicker than the thin disc).
In particular, stars born in the cold disc that migrated to the warm disc have higher iron abundances compared with particles born in the warm disc.

The results of the previous figures indicate that stellar migration between galactic components, since the birth of stars until $z=0$, is more important for the bulge and the warm disc.
In the case of the bulge, the mixing of stars born in different components -- with the exception of the stellar halo -- is responsible for its chemical patterns, but they all contribute with similar abundance distributions.
For the warm disc, a significant contribution comes from the cold disc, and therefore, the chemical patterns are also linked to the enrichment history of the cold disc\footnote{It is worth noting that the amount of stars that migrated from the cold to the warm disc might be affected by our separation between these two components; however we have checked that using different thresholds does not significantly affect our results.}.
Most stars in the cold disc were born in the cold disc of the main progenitor, which is consistent with our previous findings, indicating that the cold disc is the most homogeneous component in the Auriga sample.
Finally, the stellar halo is found to have similar contributions from stars born in the halo and in the warm/cold discs, but it is important to remark that, on average, haloes contain a considerable fraction of ex situ stars not included in this analysis.

\begin{figure*}
    \centering
    \includegraphics[draft=false]{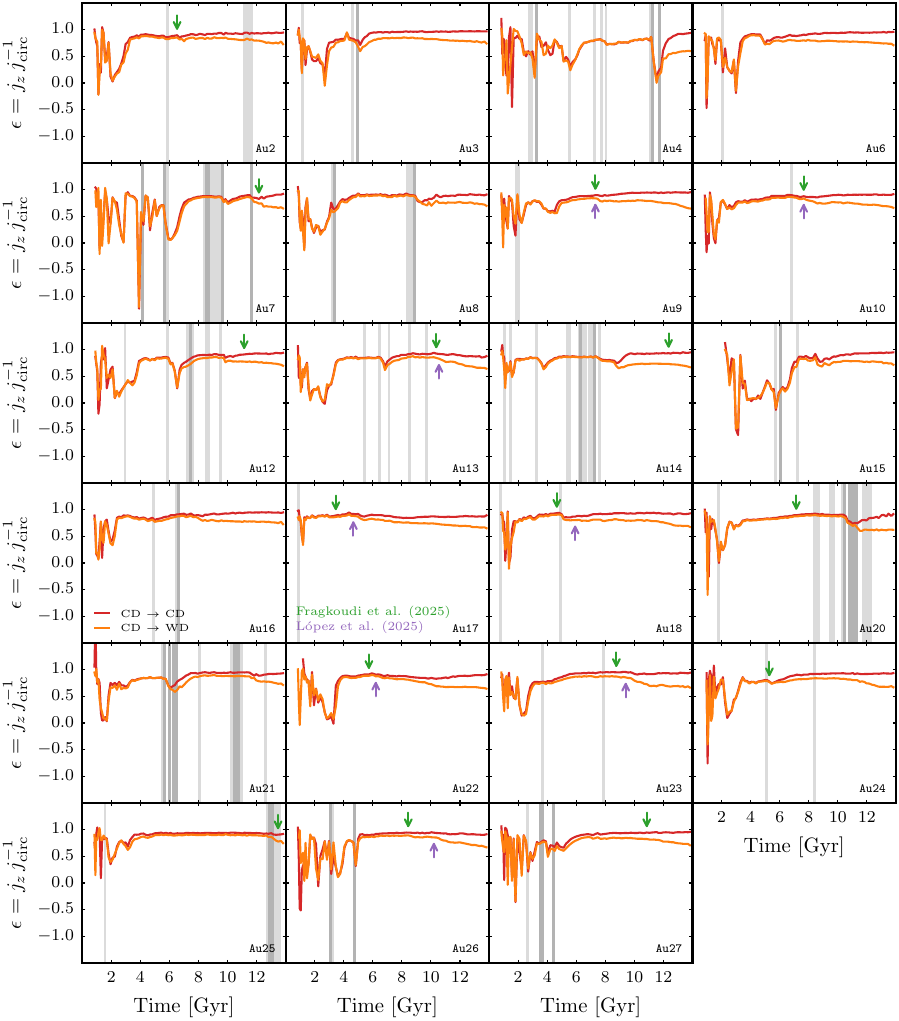}
    \caption{
    Evolution of the median circularity parameter $\epsilon$ for stars born in the cold disc that populate the warm disc at $z=0$ (orange lines) and for stars born in the cold disc that remained there until the present time (red lines). 
    The figure also shows the presence of satellites with mass ratios $\geq 3\%$ (light grey:  within $0.5 R_{200}$ and dark grey within $0.3 R_{200}$), as well as galaxies that developed bars: downward green arrows correspond to bars identified by \cite{Fragkoudi2025} and upward purple arrows to \cite{Lopez2025} which additionally developed boxy-peanut bulges.
    }
    \label{fig:warm_disc_tracking_circularity}
\end{figure*}

We now investigate in more detail the physical mechanisms driving the migration of stars from the cold disc to the warm disc discussed above.
For this, we identify stars that were born in the cold disc and are found in the cold disc at $z=0$, and compare them with stars that were born in the cold disc but migrated to the thick disc at some point during their evolution.
From Fig.~\ref{fig:warm_disc_tracking_circularity}, where we show the evolution of the median circularity parameter ($\epsilon$) for the two sub-samples, we can observe that these follow a nearly identical evolution at early times but their paths begin to diverge at varying times depending on the galaxy.
Our analysis shows that the change in $\epsilon$ found for stars that migrated from the cold to the warm disc is not due to a change in their angular momentum values, radii or height over the disc, but rather due to a variation in the orientation of the orbit, which affects the angular momentum perpendicular to the disc plane ($j_z$) and thus the $\epsilon$ values.
We found that the change in orbit orientation might be explained by the formation of a bar and/or the interaction with a satellite galaxy. In fact, bars in Auriga have been identified in 14 of the 23 galaxies of our sample; their formation times are indicated in the figure as arrows following \cite{Fragkoudi2025} and \cite{Lopez2025}\footnote{The differences seen in some cases in the formation times of the bars of these two works are due to variations in the identification of the bar region, as discussed in \cite{Lopez2025}.}. We also include shades at those times where satellites are identified near the main object (see figure caption for details).
We find that, in most cases, the decrease in the median $\epsilon$ values seen for stars moving from the cold to the warm discs correlates with the presence of a bar and/or an interaction/merger.
In the particular case of the bars, they provide a natural explanation for our findings, as stars in this component would accommodate into the different orbits possible for bars and change the $j_z$ values. Satellite interactions or mergers could also induce changes in the stellar orbits, although their effects will depend on various parameters such as mass ratio and trajectory.
While the connection between the formation of bars, mergers/interactions with satellites and internal instabilities is still a subject of debate, our results suggest that, in a cosmological context, all these mechanisms can contribute to a variation in the stellar orbits and explain our finding related to stars that are born in the cold disc and end up in the warm disc.
However, it is worth mentioning that this component migration depends on our separation into cold and warm discs, and might not be found if a different separation criterion is adopted.

\section{Conclusions}
\label{sec:conclusions}

In this work, we have investigated the chemical patterns of the various stellar components in a sample of simulated MW-mass galaxies.
To do so, we used the high-resolution cosmological simulations of the Auriga project, which consist of relatively isolated MW-mass galaxies at $z=0$ with no major mergers in the recent past.
The simulations were run with the {\sc arepo} code, a quasi-Lagrangian moving mesh code that tracks the magnetohydrodynamical evolution of gas together with collisionless dynamics for the dark matter and stellar components.

We designed a kinematic method to segregate the stars of each galaxy into four stellar components: a cold disc, a warm disc, a bulge, and a stellar halo.
This method is easily applicable to any galaxy at any time.
We have shown that our method provides an adequate kinematic and structural representation of the four components, with approximately spheroidal dispersion-dominated bulges and stellar haloes, and elongated highly rotating (warm and cold) discs.
All Auriga galaxies are disc-dominated.
On average, $\sim 70\%$ of the stellar mass is in the (cold plus warm) discs, $\sim 20\%$ is in the bulges, and the rest in the stellar haloes.
The discs and bulges have high in situ fractions ($\sim 90, 95$, and $85\%$ on average for the warm discs, cold discs, and bulges, respectively), while the stellar haloes have a high contribution from ex situ stars ($\sim 50\%$, but with significant galaxy-to-galaxy variations).

Despite some differences in the formation histories of the Auriga galaxies -- such as the occurrence and timing of mergers, differences in the accretion patterns, and variations in the formation time and growth of the disc components -- each of the four stellar components has distinct properties with systematic differences that have a determining role in their chemical history:
\begin{itemize}
    \item The haloes are always the oldest component, with a peak in mean stellar age of $\sim 10~\mathrm{Gyr}$ (the youngest stellar haloes have ages as low as $\sim 6~\mathrm{Gyr}$).
    As the majority of stars in this component are quite old, they posses low [Fe/H] and high [O/Fe] abundance ratios.
    Stellar haloes populate a specific region in the [Fe/H] versus [O/Fe] plane but show some overlap with the other galactic components, particularly with the bulge.
    \item The bulges are a more diverse population.
    While the mean stellar ages are in general in the range $8-10~\mathrm{Gyr}$, approximately half of the galaxies have a significant contribution of young stars.
    As a result, the mean [Fe/H] extends over about $1~\mathrm{dex}$ around the solar value.
    The [O/Fe] abundance ratios of the bulges vary between $0.22$ and $0.29~\mathrm{dex}$ and are, as expected, anti-correlated with the [Fe/H] ratios.
    Similar to our findings for the stellar haloes, bulges occupy a given region in the [Fe/H] versus [O/Fe] plane despite the variations among galaxies.
    \item The cold discs present the lowest mean stellar ages -- between $0$ and $8~\mathrm{Gyr}$, with a peak at $5~\mathrm{Gyr}$ -- and are the most chemically homogeneous component.
    The mean [Fe/H] ratios peak around the solar values, with a small scatter of less than $0.2~\mathrm{dex}$ and [O/Fe] values in the range $0.21$--$0.25~\mathrm{dex}$.
    These ratios, as in the case of the bulges, are clearly anti-correlated.
    \item The warm discs show values of stellar age, [Fe/H], and [O/Fe] between that of the cold discs and that of the bulges.
    Galaxy-to-galaxy variations are larger compared to those found for the cold discs but systematically smaller than those for the bulges.
\end{itemize}
These general trends are a consequence of the well-defined age-metallicity relation that we find for all galaxies in the Auriga sample.

We measured the radial gradients of the [Fe/H] abundance ratios for the cold discs and obtained a mean slope of $-0.0139~\mathrm{dex}\,\mathrm{kpc}^{-1}$.
We find that, except for the work of \cite{Lemasle2008}, the mean slope obtained does not agree with most observational results.
However, it is worth mentioning that a significant fraction of the galaxies with slopes consistent with the result of \cite{Lemasle2008} are consistent with an inside-out formation scenario.

When we looked at individual galaxies, we also found systematic trends in the abundances of the different components, except for the bulge.
In all cases, the stellar halo is the least enriched component, with a [Fe/H] abundance lower than that of the cold disc by about $0.6~\mathrm{dex}$.
The warm disc has a lower but similar abundance -- typically around $-0.1~\mathrm{dex}$ compared to the cold disc.
The relative abundance of the bulge and cold disc is more diverse, with about half of the galaxies having more enriched bulges and the other half having more enriched cold discs.
Again, this results from the stellar age distribution of the bulges, which in some cases have a significant contribution of young stars, thus affecting their $z=0$ chemical properties.

We also investigated the origin of the $z=0$ chemical patterns of the four stellar components.
For this, we compared the abundances of stars at their birth time and at $z=0$, a method that can only be applied to in situ stars.
We first showed that ex situ stars do not affect our results in any significant way because they are either a small fraction of the stellar mass of a given component (the case of bulges and discs) or the in situ and ex situ populations have similar chemical abundances (which is relevant in the case of stellar haloes, which have high ex situ fractions).
We found that stellar migration from one component to another is unimportant for the cold discs, as most stars that are in this component at the present day were formed in the cold disc of the progenitor galaxy.
We did not find significant effects on the chemical properties of the warm discs even though a significant fraction of their stars were in the cold disc at birth.
This results from the chemical similarity of these two components.
In contrast, the $z=0$ bulges are formed by stars that formed in the bulges and the (warm and cold) discs of the progenitors in similar proportions, and thus their chemical properties are affected by the enrichment history of the discs.
Finally, we found that although the stellar haloes are formed by stars born in the other stellar components, they are all old stars, have similar chemical abundances, and share similar chemical properties as those of stars that were born in situ in the stellar haloes.

Our findings show that the chemical properties of MW-mass, disc-dominated galaxies result from the relative amount of stellar mass and the enrichment histories of the different stellar components.
The enrichment histories follow a relatively simple pattern determined mainly by the age of stellar populations.
While chemical inhomogeneity -- most notably in the case of bulges -- leads to variations in the overall stellar abundances, our results show relatively narrow ranges of possible [Fe/H] and [O/Fe] ratios for the sample of galaxies at the studied mass scale and morphology.

\begin{acknowledgements}
We thank the anonymous referee for their useful comments that helped to improve this work.
CS and SEN are members of the Carrera del Investigador Científico of CONICET.
They acknowledge support from CONICET (PIBAA R73734), Agencia Nacional de Promoci\'on Cient\'{\i}fica y Tecnol\'ogica (PICT 2021-GRF-TI-00290) and UBACyT (20020170100129BA).
RJJG acknowledges support from an STFC Ernest Rutherford Fellowship (ST/W003643/1).
FAG acknowledges support from the FONDECYT Regular grant 1211370, by the ANID BASAL project FB210003, and from the HORIZON-MSCA-2021-SE-01 Research and Innovation Programme under the Marie Sklodowska-Curie grant agreement number 101086388.
\end{acknowledgements}

\bibliographystyle{aa}
\bibliography{bibliography}

\begin{appendix}

\onecolumn
\section{Galactic decomposition phase space}

\begin{figure*}[h!]
    \centering
    \includegraphics[scale=1.0]{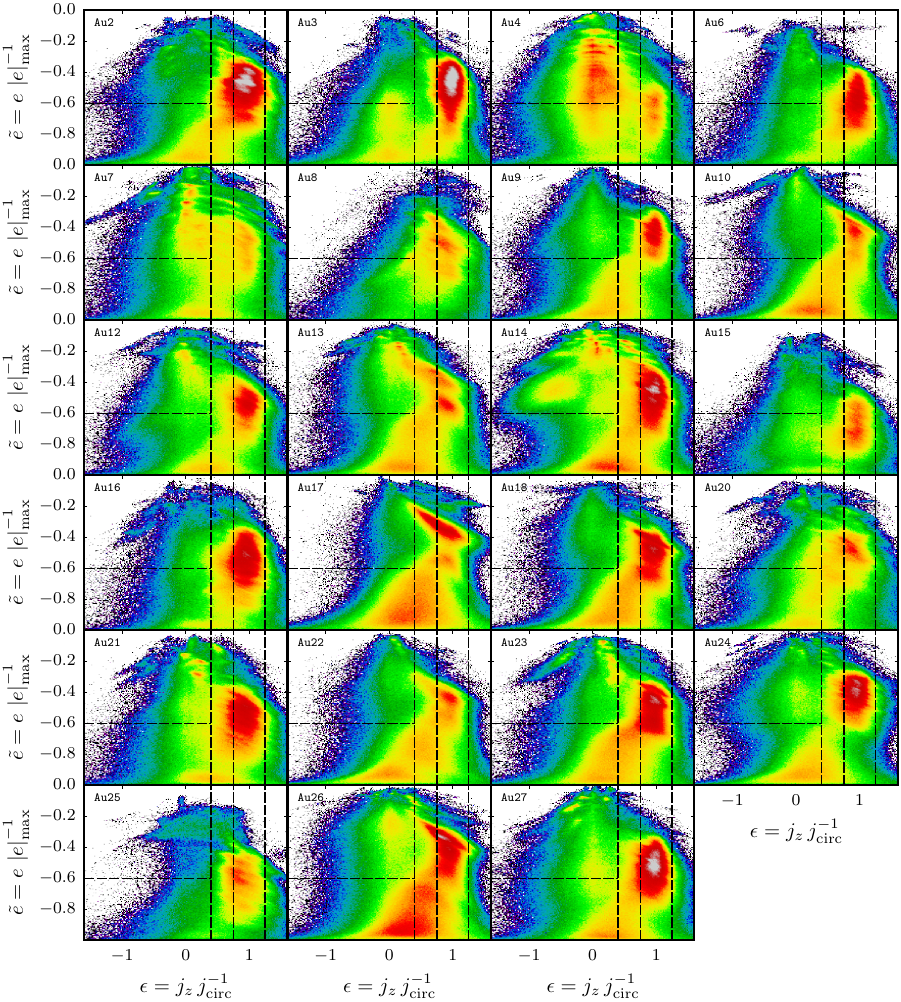}
    \caption{
    Distribution of stars in the gravitational potential-circularity parameter space for all the Auriga galaxies analysed in this work at $z=0$.
    Dashed lines indicate the separation into the different components (see also Fig.~\ref{fig:phase_space_au6}).
    }
    \label{fig:galaxy_decomposition_phase_space}
\end{figure*}

Fig.~\ref{fig:galaxy_decomposition_phase_space} shows the stellar distribution in the gravitational potential-circularity plane for all Auriga galaxies, at $z=0$.
Similarly to our findings for Au6 (Fig.~\ref{fig:phase_space_au6}), most galaxies have well-defined, rotating disc components.
All panels use the same logarithmic scale for the colour map that covers three orders of magnitude of particles per pixel.

\end{appendix}

\end{document}